\documentclass{aa}  
\usepackage{graphicx}
\usepackage{multirow}
\newcommand{\bm}[1]{{\mbox{\boldmath $#1$}}}
\usepackage{color}

\usepackage{here}

\newcommand{\ii}{{{\rm i}}}

\usepackage[colorlinks=true,citecolor=blue]{hyperref}
\usepackage{arydshln}
\usepackage{natbib}

\usepackage{comment}

\usepackage{booktabs}

\usepackage{adjustbox}

\usepackage{ulem}

\usepackage{txfonts}
\begin{document} 
   \title{
   Theory of solar oscillations in the inertial frequency range:\\
    Amplitudes of equatorial modes from a nonlinear rotating convection simulation
   }
   
   \titlerunning{Equatorial modes extracted from a solar rotating convection simulation}

   \author{Yuto Bekki
          \inst{1}
          \and
          Robert H. Cameron\inst{1}
          \and
          Laurent Gizon\inst{1,2,3}
          }


   \institute{Max-Planck-Institut f{\"u}r Sonnensystemforschung,
              Justus-von-Liebig-Weg 3, 37077 G{\"o}ttingen, Germany\\
              \email{bekki@mps.mpg.de}
         \and
         Institut f{\"u}r Astrophysik, Georg-August-Universt{\"a}t G{\"o}ttingen,
         Friedrich-Hund-Platz 1, 37077 G{\"o}ttingen, Germany
         \and
         Center for Space Science, NYUAD Institute,
         New York University Abu Dhabi, Abu Dhabi, UAE
             }

   \date{Received <-->; accepted <-->}


  \abstract
{
Several types of inertial modes have been detected on the Sun.
Properties of these inertial modes have been studied in the linear regime but have not been studied in nonlinear simulations of solar rotating convection.
Comparing the nonlinear simulations, the linear theory, and the solar observations is important to better understand the differences between the models and the real Sun. 
} 
{
We wish to detect and characterize the modes present in a nonlinear numerical simulation of solar convection, in particular to understand the amplitudes and lifetimes of the modes. 
}
{
We developed a code with a Yin-Yang grid to carry out fully-nonlinear numerical simulations of rotating convection in a spherical shell. 
The stratification is solar-like up to the top of the computational domain at $0.96\ R_\odot$. 
The simulations cover a duration of about 15 solar years, which is more than the observational length of the Solar Dynamics Observatory (SDO).
Various large-scale modes at low frequencies (comparable to the solar rotation frequency) are extracted from the simulation.
Their characteristics are compared to those from the linear model and to the observations.
}
{
Among other modes, both the equatorial Rossby modes and the columnar convective modes are seen in the simulation.
The columnar convective modes, with north-south symmetric longitudinal velocity $v_{\phi}$, contain most of the large-scale velocity power outside the tangential cylinder and substantially contribute to the heat and angular momentum transport near the equator.
Equatorial Rossby modes with no radial node ($n=0$) are also found:
They have the same spatial structures as the linear eigenfunctions.
They are stochastically excited by convection and have the amplitudes of a few m~s$^{-1}$ and mode linewidths of about 20--30 nHz, which are comparable to those observed on the Sun.
We also confirm the existence of the ``mixed Rossby modes'' between the equatorial Rossby modes with one radial node ($n=1$) and the columnar convective modes with north-south antisymmetric $v_{\phi}$ in our nonlinear simulation, as predicted by the linear eigenmode analysis.
We also see the high-latitude mode with $m=1$ in our nonlinear simulation but its amplitude is much weaker than that observed on the Sun.
}
{}


   \maketitle
%

\section{Introduction}

Large-scale convection in the Sun is still poorly understood \citep[e.g.,][]{hanasoge2012,gizon2012}.
Numerical simulations are unable to explain how thermal energy and angular momentum are transported inside the Sun's convection zone in a way that is consistent with the observations \citep[e.g.,][]{karak2018,nelson2018}.
This problem is often called the ``convective conundrum'' and is regarded as one of the most important open questions  in solar physics \citep[e.g.,][]{omara2016,brandenburg2016,hanasoge2020,vasil2021}.

Recent observations indicate that a significant component of  the large-scale non-axisymmetric solar flows  are due to  inertial modes of oscillation  \citep[][]{loeptien2018,gizon2021}.
The restoring force for these global-scale low-frequency modes of oscillation is the Coriolis force, and thus their oscillation periods are comparable to the solar rotation period ($\approx 27$ days).
In addition to the high-frequency acoustic modes, these inertial modes are expected to be useful as a tool to probe the interior of the Sun \citep[][]{gizon2021,bekki2021_l}.

The inertial modes observed and identified on the Sun include the equatorial Rossby modes, the high-latitude modes, and the critical-latitude modes
\citep[][]{loeptien2018,gizon2021,bekki2021_l,fournier2022}.
The high-frequency retrograde modes recently reported by \citet[][]{hanson2022} are also likely inertial modes.
Simplified theoretical studies have been carried out in the linear regime under the assumption of  uniform rotation \citep[][]{provost1981, saio1982, wolff1986,damiani2020} and in the case of differential rotation \citep[][]{baruteau2013,gizon2020,bekki2021_l,fournier2022}.
However, there has been no study in the fully-nonlinear regime where turbulent convection strongly interacts with these inertial modes.
Nonlinear simulations are also required in order to understand the excitation mechanism and the amplitudes of these modes.

Another interesting type of large-scale vorticity modes that might be relevant to the Sun are columnar convective modes (or ``thermal Rossby waves'').
They have been repeatedly predicted in numerical models of solar rotating convection \citep[e.g.,][]{gilman1981,glatzmaier1984,miesch2000,miesch2008} but they have not been observed on the Sun. 
A recent linear eigenmode analysis has revealed that the equatorial Rossby modes with one radial node ($n=1$) share properties with the columnar convective modes with north-south antisymmetric $z$-vorticity \citep[][]{bekki2021_l}, where $z$ is the coordinate along the rotational axis.
These so-called ``mixed Rossby modes'' have not yet been studied using  nonlinear rotating convection simulations.

In this paper, we identify and characterize properties of these low-frequency modes of oscillation in a numerical simulation of solar-like rotating convection and study how they are affected by turbulent convection.
We extract these modes from a fully nonlinear simulation and compare their mode properties such as dispersion relations and eigenfunctions with those of the linear eigenmodes reported by \citet[][]{bekki2021_l}.
In addition, we look at mode amplitudes which cannot be discussed in the linear regime.

The organization of the paper is as follows.
In \S \ref{sec:inertialmodes}, we shortly review  previous studies on the various types of solar inertial modes.
In \S \ref{sec:model}, our numerical model is explained in detail.
We also describe the analysis method for extracting the global-scale modes of oscillation from a temporal series of simulation data.
We report the extracted columnar convective modes in \S \ref{sec:thRos}, and the equatorial Rossby modes in \S \ref{sec:eqRos}.
The newly-discovered mixed Rossby mode is presented in \S \ref{sec:mixed}.
In all cases, the extracted modes are compared with the results of linear eigenmode analysis.
The transport properties by these modes are discussed in \S \ref{sec:transport}.
Finally, possible implications are discussed with concluding remarks in \S \ref{sec:summary}.

\section{Inertial modes on the Sun} \label{sec:inertialmodes}

In this paper we will primarily focus on the equatorial Rossby modes (\S~\ref{sec:intro_rossby}), the columnar convective modes (\S~\ref{sec:intro_thermal}), and the ``mixed Rossby modes'' (\S~\ref{sec:intro_mixed}).
The other inertial modes discussed in \S~\ref{sec:intro_other} are beyond the scope of the current paper.

\subsection{Equatorial Rossby modes} \label{sec:intro_rossby}

The classical Rossby modes \citep[][]{rossby1939,rossby1940} are modes of radial vorticity originating from the planetary $\beta$-effect.
In the case of a uniformly-rotating sphere, these modes propagate in the retrograde direction (opposite to rotation) in a co-rotating frame.
They are essentially incompressible modes and the associated motion is quasi-toroidal.
In the solar and stellar context, they are also commonly known as r modes \citep[e.g.,][]{papaliozou1978,saio1982}.

The existence of these Rossby modes on the Sun is well established \citep{loeptien2018,hanasoge2019,liang2019,proxauf2020,mandal2020,hanson2020,hathaway2020,gizon2021,mandal2021}.
They are observed for azimuthal orders in the range $3\leq m \leq 15$ and follow the dispersion relation of the classical sectoral ($l=m$) Rossby modes, where $m$ is the azimuthal order and $l$ is the spherical degree.
It is found that these equatorial Rossby modes contribute a significant fraction of the large-scale horizontal velocity power at low latitudes.

\subsection{Columnar convective modes} \label{sec:intro_thermal}

Another type of modes with large-scale vorticity, which might be relevant to the Sun, are the columnar convective modes. 
These modes are prograde-propagating convective columns that are strongly rotationally-constrained and are thus aligned parallel to the rotation axis \citep[e.g.,][]{unno1989}.
They are also known as ``thermal Rossby waves'' particularly among the geophysical fluid dynamics community because they are thermally (convectively) driven and they result from the conservation of potential vorticity \citep{busse1970,busse2002}.
When the term ``thermal Rossby waves'' was first introduced by \citet{busse1986b}, an incompressible fluid was considered and thus the propagation frequency of these convective modes is purely set by the ``topographic $\beta$ effect'' originating from the curvature of the spherical boundaries.
However, when it comes to highly-stratified compressible fluids such as the interiors of the Sun and stars, there is an additional $\beta$ effect, namely, the ``compressional $\beta$ effect'' originating from the strong background density stratification \citep[][]{glatzmaier1981,ingersoll1982,evonuk2008,glatzmaier2009,evonuk2012,verhoeven2014,ong2020}.
In the solar and stellar physics community, the term ``thermal Rossby waves'' has been sometimes used to describe the prograde-propagating convective columns whose propagation frequencies are in fact affected by both the topographic and compressional $\beta$ effects \citep{miesch2008}.
In order to avoid this ambiguity, we will follow the convention of \citet[][]{bekki2021_l} and will primarily use the term ``columnar convective modes'' in this paper.

In numerical simulations of solar-like rotating convection simulations, the columnar convective modes can be seen as north-south aligned downflow lanes across the equator at the surface \citep[e.g.,][]{miesch2008,bessolaz2011,matilsky2020}.
They are often called ``banana cells'' and are regarded as the most efficient convective structure in terms of the thermal energy transport under the rotational constraint \citep[e.g.,][]{miesch2000,brun2004,miesch2008,kapyla2011,gastine2013,hotta2014b,featherstone2016,hindman2020}.
Furthermore, they are also believed to play a significant role in transporting the angular momentum equatorward to maintain the differential rotation \citep[e.g.,][]{gilman1986,miesch2000,balbus2009,matilsky2020}.
However, despite their significance, the columnar convective modes have never been successfully detected on the Sun. 
It still remains unclear whether they really exist in the deep convection zone and are not visible at the surface for some reason, or if they are simply absent in the Sun.

The dispersion relation and eigenfunctions of the columnar convective modes were first derived by \citet{glatzmaier1981} using a one-dimensional cylinder model, and recently, by \citet{bekki2021_l} using a more realistic two-dimensional model of the solar convection zone.
A local analysis of these modes has also been carried out by \citet[][]{hindman2022} in the context of low-mass stars.
As far as the authors recognize, however, there is no such a study that compares the linear dispersion relation and eigenfunctions of the columnar convective modes with those found in a fully-nonlinear simulation of rotating convection of the Sun.

\subsection{Mixed Rossby modes} \label{sec:intro_mixed}

Using the linear model of solar inertial oscillations, \citet{bekki2021_l} have recently found that the equatorial Rossby modes with one radial node ($n=1$) and the columnar convective modes with north-south antisymmetric $z$-vorticity $\zeta_{z}$ are mixed with each other.
This newly-discovered ``mixed Rossby mode'' has a dispersion that asymptotically (at large azimuthal wavenumbers $m$) approaches to that of the equatorial Rossby mode with no radial node ($n=0$) for the retrograde-propagating part ($\omega<0$) and to that of the north-south $\zeta_{z}$-symmetric columnar convective modes for the prograde-propagating part ($\omega>0$).
Here, $\omega$ is the frequency measured in the rotating frame.
Due to this mode mixing, the $n=1$ equatorial Rossby modes (retrograde-propagating ``mixed Rossby modes'') become partially convective and have substantial radial motions. 
This is in contrast to the $n=0$ modes where fluid motions are quasi-toroidal.
Therefore, it is expected that the ``mixed Rossby modes'' might play a role in transporting thermal energy and angular momentum in the Sun's convection zone.

\subsection{Other inertial modes} \label{sec:intro_other}

At higher latitudes (above $60^\circ$), modes with $1\leq m \leq 4$
were also detected, which were shown to be global modes of inertial oscillation of the full convection zone. 
In particular, the $m=1$ high-latitude mode can be observed at all latitudes (with much smaller amplitude) with the exact same frequency. 
This $m=1$ mode has the largest amplitude and is associated with a spiralling flow pattern in the longitudinal velocity around the poles. 
It is the explanation for the observations reported by \citet{hathaway2013} and \citet{bogart2015}. 
The dispersion relation and the eigenfunctions of the high-latitude modes can be well reproduced by a linear analysis of inertial oscillations in a realistic solar convection zone model \citep{bekki2021_l}.

At middle latitudes, additional retrograde inertial modes with $m\leq 10$ have been observed near their critical latitudes, where the mode angular frequencies are equal to the differential rotation rate \citep[][]{gizon2021}. 
Some of these mid-latitude modes have been identified in the linear model of \citet{bekki2021_l}, see e.g. the $m=2$ mode reported by \citet{gizon2021}. 
A one-dimensional linear analysis of modes on differentially rotating spheres by \citet{fournier2022} also predicts such critical-latitude modes to be present.

Furthermore, \citet{hanson2022} have recently reported additional low-amplitude modes of north-south antisymmetric vorticity near the equator that propagate in a retrograde direction with $8\leq m \leq 14$. 
These modes are also likely inertial modes, according to the simplified linear analysis by \citet{triana2022}. 
Whether these modes are also present in the more realistic solar models remains to be studied.

\section{Methods} \label{sec:model}

\subsection{Numerical model of rotating convection} \label{sec:modeleq}

We have developed a code to solve three-dimensional fully-compressible hydrodynamic equations in a rotating spherical shell.
With the reduced-speed of sound approximation, the hydrodynamic equations are expressed in a spherical coordinate $(r,\theta,\phi)$ as
\citep[e.g.,][]{hotta2014a}:
\begin{eqnarray}
&&\frac{\partial \rho_{1}}{\partial t}=-\frac{1}{\xi^{2}}\nabla\cdot(\rho_{0}\bm{v}), \label{eq:continuity} \\
&&\frac{\partial\bm{v}}{\partial t}=-\bm{v}\cdot\nabla\bm{v}-\frac{\nabla p_{1}}{\rho_{0}}-\frac{\rho_{1}}{\rho_{0}}g\bm{e}_{r}
	+2\bm{v}\times\Omega_{0}+\frac{1}{\rho_{0}}\nabla\cdot\bm{\mathcal{D}}, \label{eq:motion} \\
&&\frac{\partial s_{1}}{\partial t}=-\bm{v}\cdot \nabla s_{1}+\frac{1}{\rho_{0}T_{0}} \nabla\cdot (\rho_{0}T_{0}\kappa\nabla s_{1}) \nonumber \\
&& \ \ \ \ \ \ \ \ \ \ \ +\frac{1}{\rho_{0}T_{0}} (\bm{\mathcal{D}}\cdot\nabla)\cdot\bm{v}
	+\frac{1}{\rho_{0}T_{0}}(Q_{\mathrm{heat}}+Q_{\mathrm{cool}}), \label{eq:entropy}  \\
&&\frac{p_{1}}{p_{0}}=\gamma\frac{\rho_{1}}{\rho_{0}}+\frac{s_{1}}{c_{\mathrm{v}}}. \label{eq:state}
\end{eqnarray}
Here, $\xi=100$ denotes a factor by which the background sound speed is reduced to relax the severe CFL condition \citep{rempel2005,hotta2014a}.
The quantities with subscript $0$, $p_{0}$, $\rho_{0}$, $T_{0}$, and $g_{0}$, represent the pressure, density, temperature, and gravitational acceleration of the time-independent background which is in an adiabatically-stratified hydrostatic equilibrium.
Also, $\Omega_{0}$ denotes the rotation rate of the rigidly-rotating radiative core and we use the solar value of $\Omega_{0}/2\pi=431.3$ nHz.
We use the same solar-like background stratification model as \citet{rempel2005} and \citet{bekki2017a} from $r_{\mathrm{min}}=0.71R_{\odot}$ to $r_{\mathrm{max}}=0.96R_{\odot}$ where $R_{\odot}$ is the solar radius.
The variables $v$, $\rho_{1}$, $p_{1}$, and $s_1$ represent perturbations from the background reference state. 
The equations are fully nonlinear, however the Eq.~(\ref{eq:state}) assumes that these perturbations are not too large to prevent us from linearizing the equation of state.
As is usual, $c_v$ denotes the specific heat at constant volume and the ratio of specific heats is given by $\gamma = 5/3$.

The viscous stress tensor, $\bm{\mathcal{D}}$, is given by
\begin{eqnarray}
&& \mathcal{D}_{ij}=\rho_{0}\nu\left[ \mathcal{S}_{ij}-\frac{2}{3}(\nabla\cdot\bm{v})\delta_{ij}\right],
\end{eqnarray}
where $\mathcal{S}$ is the deformation tensor. 
See \citet{fan2014} for the expression of this tensor in spherical coordinates.
The coefficients $\nu$ and $\kappa$ are respectively the eddy viscosity and the eddy  thermal diffusivity, which model the unresolved subgrid-scale turbulent motions.
In this study, we use the spatially-uniform turbulent viscosity $\nu=10^{12}$ cm$^{2}$~s$^{-1}$ and omit thermal diffusivity $\kappa=0$.
This enhances the effective Prandtl number and thus mimics the highly-magnetized convection \citep{hotta2015,bekki2017b}.

The internal heating and cooling terms, $Q_{\mathrm{heat}}$ and $Q_{\mathrm{cool}}$, are specified similarly to \citet{karak2018}:
The radiative heating is assumed to be proportional to the difference of the background pressure from its surface value,
\begin{eqnarray}
&& Q_{\mathrm{heat}}=\alpha\left[ p_0(r)-p_0(r_{\mathrm{max}})\right],
\end{eqnarray}
where the normalization factor $\alpha$ is determined so that
\begin{eqnarray}
&& L_{*}=4\pi\int_{r_{\mathrm{min}}}^{r_{\mathrm{max}}}r^{2}Q_{\mathrm{heat}}(r)dr,
\end{eqnarray}
where $L_{*}$ is the luminosity.
The functional form of $Q_{\mathrm{heat}}$ gives a good approximation of the radiative flux computed from the solar temperature and opacity values of Model S \citep{featherstone2016}.
The radiative cooling at the surface is assumed to have a thickness comparable to the local pressure scale height $H_{p}$, and thus is given by
\begin{eqnarray}
&&Q_{\mathrm{cool}}=-\frac{1}{r^{2}}\frac{\partial}{\partial r}(r^{2} F_{\mathrm{sf}}),
\end{eqnarray}
where the surface cooling flux $F_{\mathrm{sf}}$ is specified as
\begin{eqnarray}
&&F_{\mathrm{sf}}=\frac{L_{*}}{4\pi r^{2}}\exp{\left[-\left( \frac{r-r_{\mathrm{max}}}{H_{p}(r_{\mathrm{max}})}\right)^{2} \right]}.
\end{eqnarray}
In this study, we reduce the luminosity from the solar value $L_{\odot}=3.84\times 10^{33}$ erg s$^{-1}$ by a factor of 20, i.e., $L_{*}=L_{\odot}/20$.
By doing so, we reduce the convective Rossby number $\mathrm{Ro}$ ($\propto L_{*}^{1/3}$), which helps to produce a solar-like differential rotation (with a faster equator and slower poles) in a rotating convection simulation \citep{gastine2013,fan2014,hotta2014b,kapyla2014}.

\begin{figure}
\begin{center}
\includegraphics[width=\linewidth]{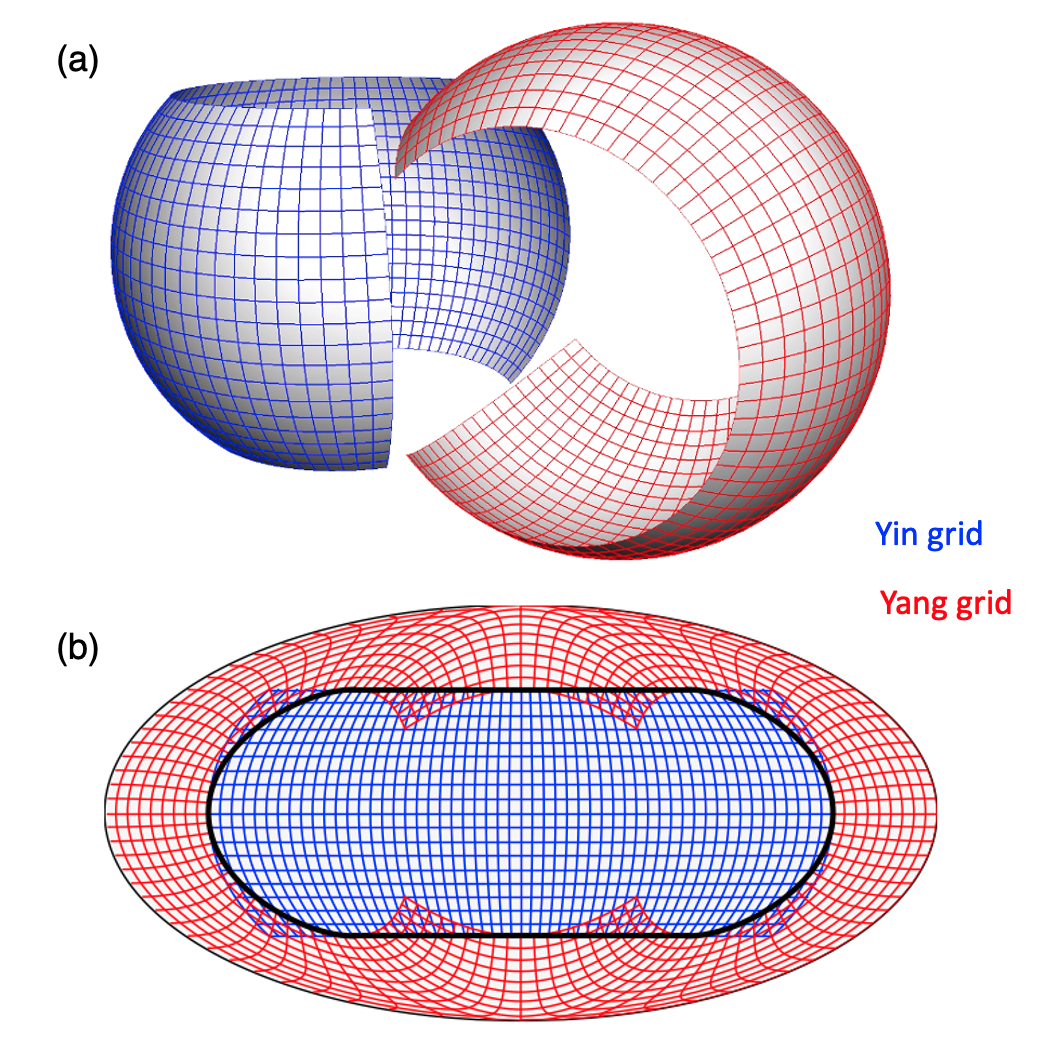}
\caption{
Yin-Yang grid used in our simulation. 
(a) Three-dimensional view of the Yin and Yang grids.
Blue and red lines show Yin and Yang coordinates, respectively. 
(b) Mollweide projection of the Yin-Yang grid.
The black thick curve denotes the location where the horizontal boundary condition is set in our code.
}
\label{fig:yinyang}
\end{center}
\end{figure}

We solve the Eqs.~(\ref{eq:continuity})--(\ref{eq:entropy}) using a fourth-order centered-differencing method for spatial derivatives and a four-step Runge-Kutta scheme for the time integration \citep[e.g.,][]{voegler2005}. 
To minimize numerical artifacts while allowing us to operate at as low a thermal diffusivity as possible, we use the slope-limited artificial diffusion presented in \citet{rempel2014} for entropy $s_{1}$. 
In order to compute in a full-spherical shell while avoiding the coordinate singularity at the poles, a Yin-Yang grid is adopted \citep{kageyama2004}.
The Yin and Yang grids are defined so that they cover a full spherical surface in a way shown in Fig.~\ref{fig:yinyang}.
The grids extend in latitudes ($\pi/4<\theta<3\pi/4$) and longitudes ($-3\pi/4<\phi<3\pi/4$), respectively.
However, unlike the method proposed in \citet{kageyama2004}, we set the boundary condition on the curve $\mathcal{C}$,
\begin{eqnarray}
&& \mathcal{C}:-\pi<\phi<\pi,  \ \theta=\pi/4,3\pi/4, 
\end{eqnarray}
and on the curve $\mathcal{C}'$,
\begin{eqnarray}
&& \mathcal{C}':\left\{
    \begin{array}{l}
      \theta '=\cos^{-1}{\left[\sin{\theta  \sin{\phi }}\right]}  \\
      \phi '=\tan^{-1}{\left[ -\cos{\theta } / (\sin{\theta }\cos{\phi })\right]}
    \end{array}
  \right.
\end{eqnarray}
where $(\theta ,\phi)\in \mathcal{C}$.
By doing so, the overlapping regions are excluded from our numerical domain.
The location where the boundary condition is set on a spherical surface is shown in a thick black curve in Fig.~\ref{fig:yinyang}b.
Both the upper and lower radial boundaries are assumed to be impenetrable and stress-free, and the radial gradient of entropy is assumed to vanish there.
The grid resolution is $72(N_{r})\times96(N_{\theta})\times288(N_{\phi})\times2$(Yin and Yang).
The simulation is initiated from a small random fluctuation in $s_{1}$.
To check whether the results are sensitive to the initial perturbations, we carry out 6 different simulation runs with different random initial fluctuations.
Each simulation run corresponds to about $25$ solar years, and we analyze the  $15$ years of data  after the differential rotation becomes statistically stationary, see Appendix~\ref{appendix:A}.
The data is saved at a time cadence of about $4.7$ days.
Most results shown in the following sections are averages over 6 realizations (of 15 years each) to improve the signal-to-noise ratio.


\subsection{Extracting modes from simulations} \label{sec:svd}

We begin with the simulations where each physical variable is only given for discrete values of $r$, $\theta$, $\phi$, and $t$.
We extract the eigenfunctions of the large-scale low-frequency modes of oscillation from the simulation data using a singular-value decomposition (SVD) similar to what was done by \citet{proxauf2020}.
To sketch the method, we consider $q_{\alpha}(r,\theta,\phi,t)$ to be any of the physical variables,  $v_{r}$, $v_{\theta}$, $v_{\phi}$, $s_{1}$, or $p_{1}$. 
We Fourier transform these variables 
to obtain 
\begin{eqnarray}
&& \tilde{q}_{\alpha} (r,\theta, m, \omega) = \int \  q_{\alpha}(r,\theta,\phi,t) \ e^{\ii (\omega t - m\phi)} \ dt\ d\phi ,
\end{eqnarray}
 where $m$ is the azimuthal order and $\omega$ is the angular frequency.
With this definition, the phase speed $\omega/m$ is positive  in the direction of rotation (prograde) and it is negative in the direction opposite to rotation (retrograde). 
In the following, we choose $m$ to be positive with no loss of generality.
Each $m$ is analyzed independently.

Among the set of variables $\{ q_\alpha \}$,  we choose a particular physical variable $q_{\beta}$ to target a particular mode. 
For example, we choose $q_\beta=u_\phi$ for the columnar convective modes  and $q_\beta=u_\theta$ for the equatorial Rossby modes and the ``mixed Rossby modes''.
Since our main focus is on the modes that peak near the equator, we consider the latitudinal average
\begin{eqnarray}
&& \tilde{q}_{\beta,{\rm eq}}(r,m,\omega)=
\frac{6}{\pi}\int_{\pi/2-\pi/12}^{\pi/2+\pi/12} \tilde{q}_{\beta}(r,\theta,m,\omega)\ d\theta
\end{eqnarray}
 over a narrow band of latitudes covering $15^\circ$ on either side of the equator.
Given the mode frequency, $\omega_{\rm mode}$, for which we want to extract the eigenfunctions, we limit the domain of analysis to the frequency range $[\omega_{1},\omega_{2}] \ni \omega_{\rm mode}$ and to an appropriate radius range $[r_{1},r_{2}]$ in which the mode has significant power, in order to reduce the contamination from the neighboring  modes.
For each fixed $m$, the quantity $\tilde{q}_{\beta,\mathrm{eq}}$ is then decomposed according to the SVD as
\begin{eqnarray}
&& \tilde{q}_{\beta,{\rm eq}}(r,m,\omega )=\sum_{k} \sigma^{\beta}_{k}(m) U^{\beta}_{k}(r,m) V^{\beta,H}_{k}(m,\omega),
\end{eqnarray}
where the $\sigma_{k}$ are the singular values, $U_{k}$ and $V_{k}$ are the left and right singular vectors, and $H$ denotes the conjugate transpose.
The vectors $V_{k}$ are normalized such that $V^{H}_{k} V_{k^{\prime}} = \delta_{k k^{\prime}}$.
The decomposition is ordered such that the first singular value is dominant over the other values. 
For each mode, we keep only the first of the right singular vectors, $V_{0}$, from the SVD. 
Using $V_0^\beta$ derived from $q_{\beta}$, the spatial dependence of a mode is calculated for all the other variables $q_\alpha$ according to
\begin{eqnarray}
&& q_{\alpha,\mathrm{mode}}(r,\theta,m)=\sum_{\omega^{\prime}=\omega_{1}}^{\omega_{2}} \tilde{q}_{\alpha}(r,\theta, m,\omega^{\prime})V^{\beta}_{0} (m,\omega^{\prime}).
\label{eqn:V0}
\end{eqnarray}
These spatial functions are approximations to a mode's eigenfunctions, and can be compared to the eigenfunctions from the linear analysis.
The amplitude of a mode extracted using the above equation is an estimate of the root-mean-square (rms) of this mode in 
the frequency range $\omega_1\leq \omega\leq \omega_{2}$, according to the Parseval's theorem.


\subsection{Linear eigenvalue solver}

\begin{figure}
\begin{center}
\includegraphics[width=\linewidth]{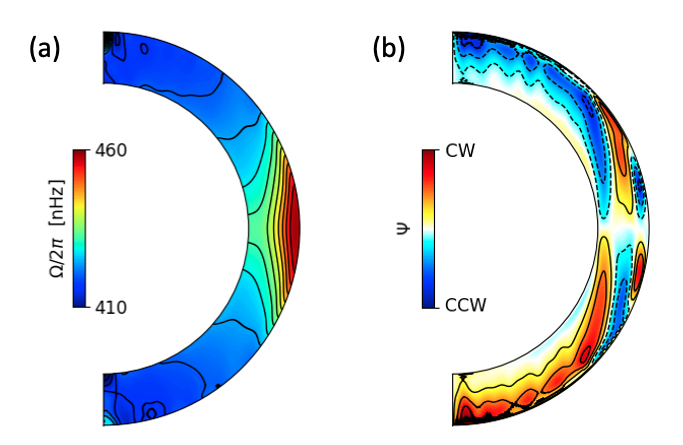}
\caption{
    Temporally-averaged profile of (a) the differential rotation $\Omega(r,\theta)=\Omega_{0}+\langle v_{\phi} \rangle/r\sin{\theta}$, and 
    (b) the streamlines of the meridional circulation $\bm{v}_{\mathrm{m}}=(\langle v_{r}\rangle,\langle v_{\theta}\rangle)$. 
    Here, $\langle \rangle$ denotes the longitudinal average.
    The meridional flow stream function $\Psi$ is defined by $\rho_{0}\bm{v}_{\mathrm{m}}=\nabla\times(\Psi\bm{e}_{\phi})$. 
    The Red (blue) indicates the circulation is clockwise (counter-clockwise), i.e., the flow is poleward near the surface at high latitudes in both hemispheres.
}
\label{fig:drmc}
\end{center}
\end{figure}

The modes extracted from the nonlinear rotating convection simulation will be compared to the linear eigenmodes of oscillation in the Sun.
For solving the linearized problem, we use the code developed in \citet[][]{bekki2021_l}.
The differences from the \citet[][]{bekki2021_l}'s setup are follows:
The lower and upper boundaries are changed to $(r_{\mathrm{min}},r_{\mathrm{max}})=(0.71 R_{\odot},0.96 R_{\odot})$ corresponding to those of the nonlinear simulation.
We also impose the differential rotation (the axisymmetric background mean flow) taken from the nonlinear simulation (Fig.~\ref{fig:drmc}a).
However, we do not take into account the meridional circulation which has a much smaller impact on inertial modes than the differential rotation \citep[][]{gizon2020,fournier2022}.
For simplicity, the background is adiabatic ($\delta=0$) and spatially-uniform viscosity of $10^{12}$ cm$^{2}$~s$^{-1}$ is included.

\section{General Results}

\begin{figure*}
\begin{center}
\includegraphics[width=0.985\linewidth]{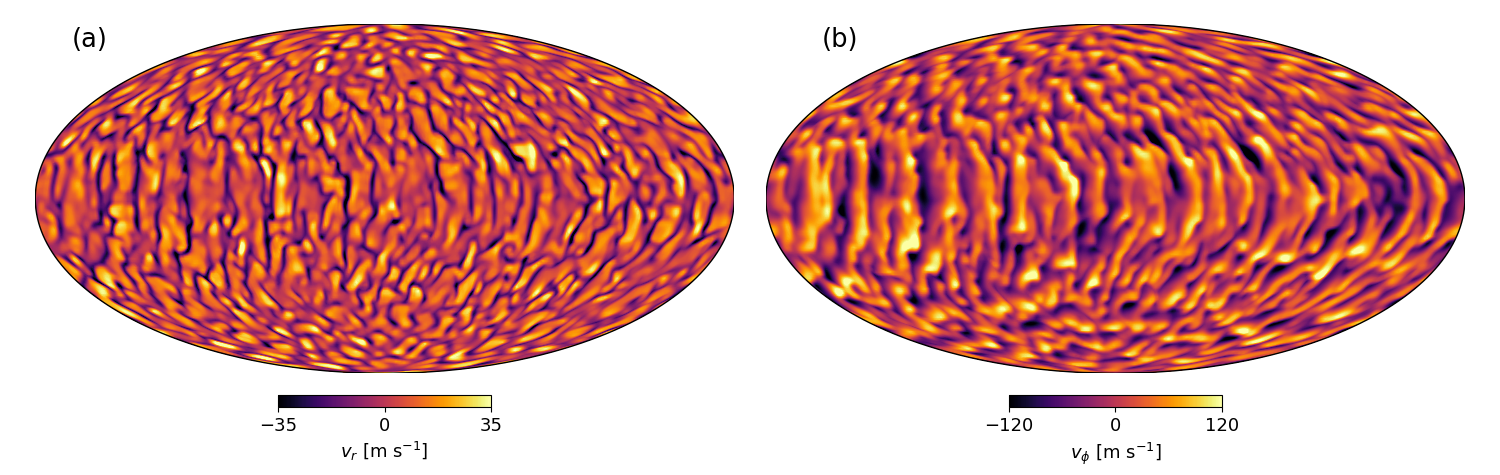}
\caption{
Snapshots of the convective pattern in our simulation near the top boundary $r=0.95R_{\odot}$.
Panels (a) and (b) show the radial velocity $v_{r}$ and non-axisymmetric component of the longitudinal velocity $v_{\phi}$.
}
\label{fig:vrs1}
\end{center}
\end{figure*}

\subsection{Rossby number regime}

Let us first evaluate the parameter regime of our nonlinear rotating convection simulation.
To do so, we compute the volume-averaged rms velocity in the simulation (fluctuations with respect to the mean flows) defined by
\begin{equation}
\overline{v}^2_{\mathrm{rms}}= \frac{1}{V} \int_{V} \ \left[  (v_{r}-\langle v_{r} \rangle)^{2} + (v_{\theta}-\langle v_{\theta} \rangle)^{2}+ (v_{\phi}-\langle v_{\phi} \rangle)^{2} \right] dV ,
\end{equation}
where $\langle \rangle$ denotes the azimuthal average and the integral is taken over the volume of the whole convection zone, $V$.
We obtain $\overline{v}_{\mathrm{rms}}=37.1$ m~s$^{-1}$, which is smaller than that of previous simulations of solar global convection by a factor of about $3$ \citep[e.g.,][]{miesch2008,hotta2022}.
This is due to the fact that the luminosity is reduced by a factor of $20$ from the solar value in our simulation.

The rotational influence on convection can be measured by the Rossby number 
\begin{eqnarray}
&& \mathrm{Ro}=\frac{\overline{v}_{\mathrm{rms}}}{2\Omega_{0} (r_{\mathrm{max}}-r_{\mathrm{min}})}.
\end{eqnarray}
We obtain $\mathrm{Ro}=0.04$, indicating that our simulation is operating in a strongly rotationally-constrained regime.
Thanks to this low $\mathrm{Ro}$, we successfully obtain the solar-like differential rotation (with faster equator and slower poles); see  \citet{gastine2013}.
Whether the Rossby number in our simulation takes a realistic value is an open question:
The Rossby number in the Sun's convection zone is one of the most important unknown global parameters, which bears on the solar convective conundrum.

\subsection{Axisymmetric mean flows}

Figures~\ref{fig:drmc}a and b show the time-averaged profiles of the axisymmetric mean flows, i.e., differential rotation and meridional circulation, respectively.
For the parameters we used, the differential rotation is solar-like although its amplitude is much weaker than that of the real Sun \citep[e.g.,][]{howe2009}.
The reference rotation rate of $\Omega_{0}/2\pi=431.3$ nHz roughly corresponds to the middle-latitude rotation rate ($\approx 40^{\circ}$) at the surface.
The meridional flow tends to be multiple-cellular structure at low latitudes but is largely counter-clockwise (clockwise) in the northern (southern) hemisphere at higher latitudes:
The flow at high latitudes is poleward (equatorward) at the surface (base of the convection zone) \citep[e.g.,][]{featherstone2015}.


\section{Low-frequency modes found in our simulation} \label{sec:modes}

\subsection{Columnar convective modes} \label{sec:thRos}

Figures~\ref{fig:vrs1}a and b show temporal snapshots of the non-axisymmetric components of radial velocity $v_{r}$ and longitudinal velocity $v_{\phi}$ near the surface $r=0.95R_{\odot}$, respectively.
The convective structure at high latitudes can be characterized by granular cells consisting of broad upflows and narrow downflows \citep[e.g.,][]{Spruit1990}.
Near the equator, we can clearly see the north-south aligned lanes of radial and longitudinal velocities.
They are often called ``banana cells'' and have been repeatedly reported in previous numerical simulations of rotating convection \citep{miesch2000,kapyla2011,gastine2013,guerrero2013,hotta2014b,featherstone2016,kapyla2019,matilsky2019,matilsky2020}.
We will show that these ``banana-cell'' features can be identified as the columnar convective modes largely originating from the compressional $\beta$-effect \citep[][]{glatzmaier1981,glatzmaier2009,verhoeven2014,bekki2021_l}.

\begin{figure*}[h]
\begin{center}
\includegraphics[width=0.99\linewidth]{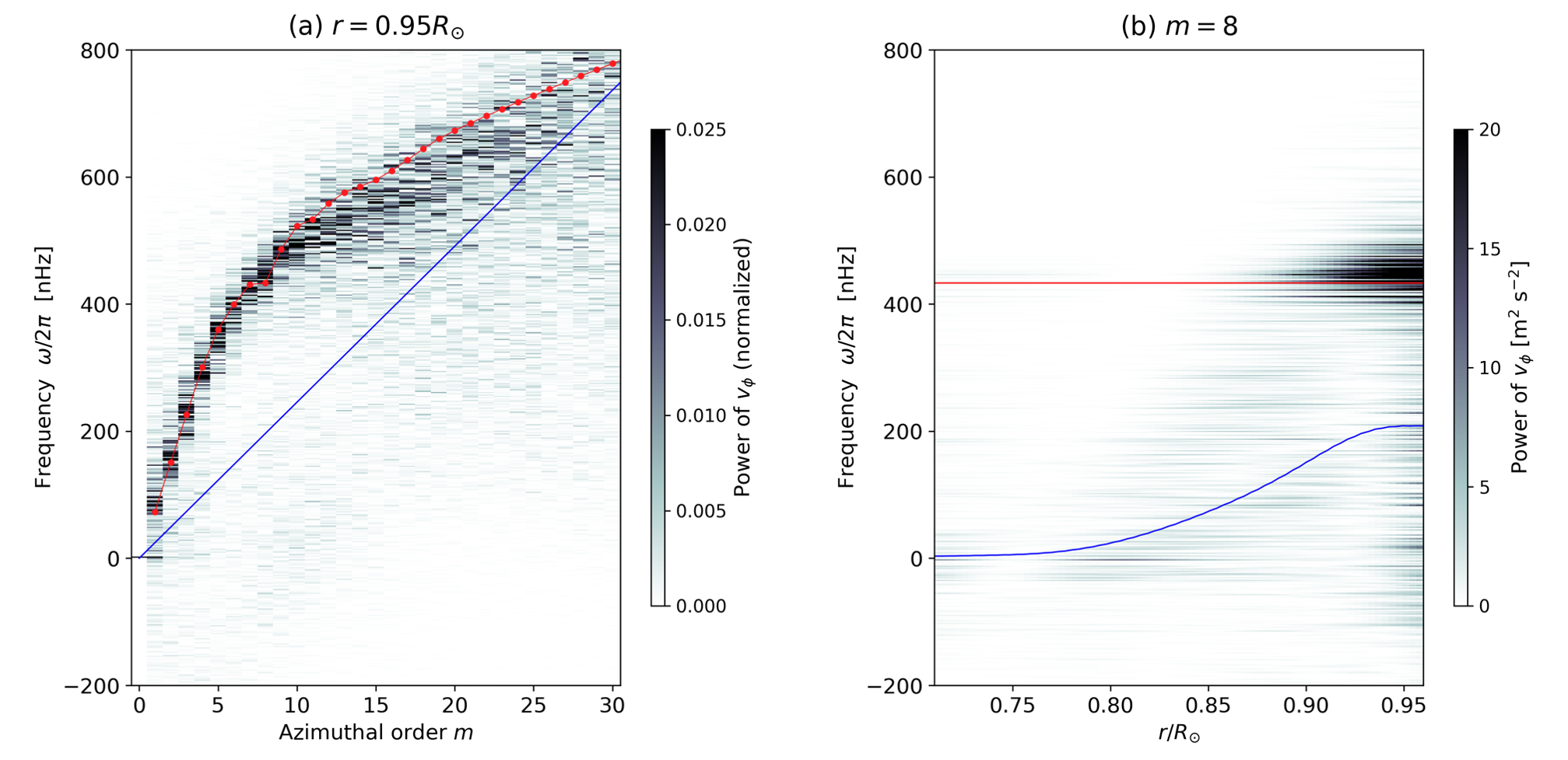}
\caption{
(a) Power spectrum of longitudinal velocity $v_{\phi}$ near the top boundary $r=0.95R_{\odot}$, averaged over the equatorial band ($\pm 15$ deg).
The power is normalized at each $m$.
The spectrum is computed in a frame rotating at $\Omega_{0}/2\pi=431.3$ nHz (the rotation rate of the radiative interior).
The blue line represents the advective speed by the local differential rotation, $m\left[\Omega(r,\pi/2)-\Omega_{0} \right]$.
Overplotted in red represents the dispersion relation of the columnar convective modes from the linear eigenmode calculation. 
Panel (b) shows the same equatorial power spectrum at fixed azimuthal order $m=8$ as a function of depth. 
The blue line is again the local advection frequency and the red line is the eigenfrequency from the linear analysis.
}
\label{fig:power_thRos}
\end{center}
\end{figure*}

\begin{figure*}[h]
\begin{center}
\includegraphics[width=0.96\linewidth]{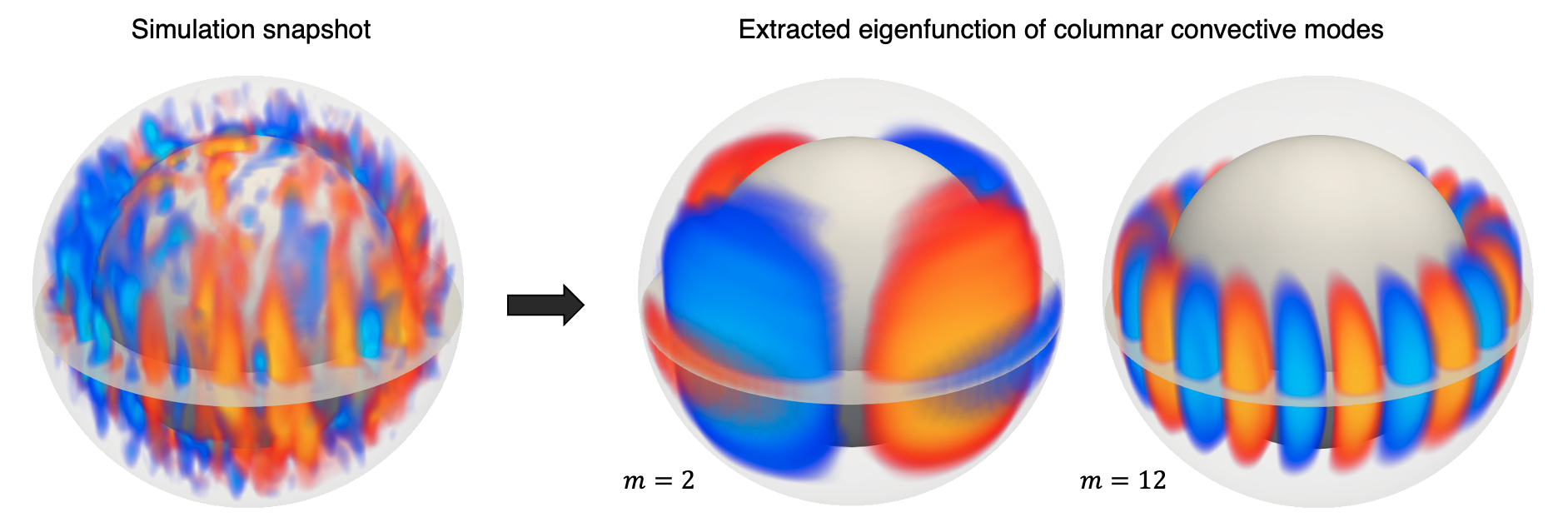}
\caption{
(Left) A snapshot of the pressure perturbation (non-axisymmetric component) from the nonlinear simulation shown as a 3D volume rendering.
Red/yellow and blue/cyan parts correspond to the regions with positive and negative pressure perturbations, respectively.
(Right) Eigenfunctions of pressure perturbation of the conlumnar convective modes extracted from the simulation data using SVD.
The cases with $m=2$ and $m=12$ are shown.
}
\label{fig:p1_thRos}
\end{center}
\end{figure*}

\begin{figure*}[]
\begin{center}
\includegraphics[width=0.95\linewidth]{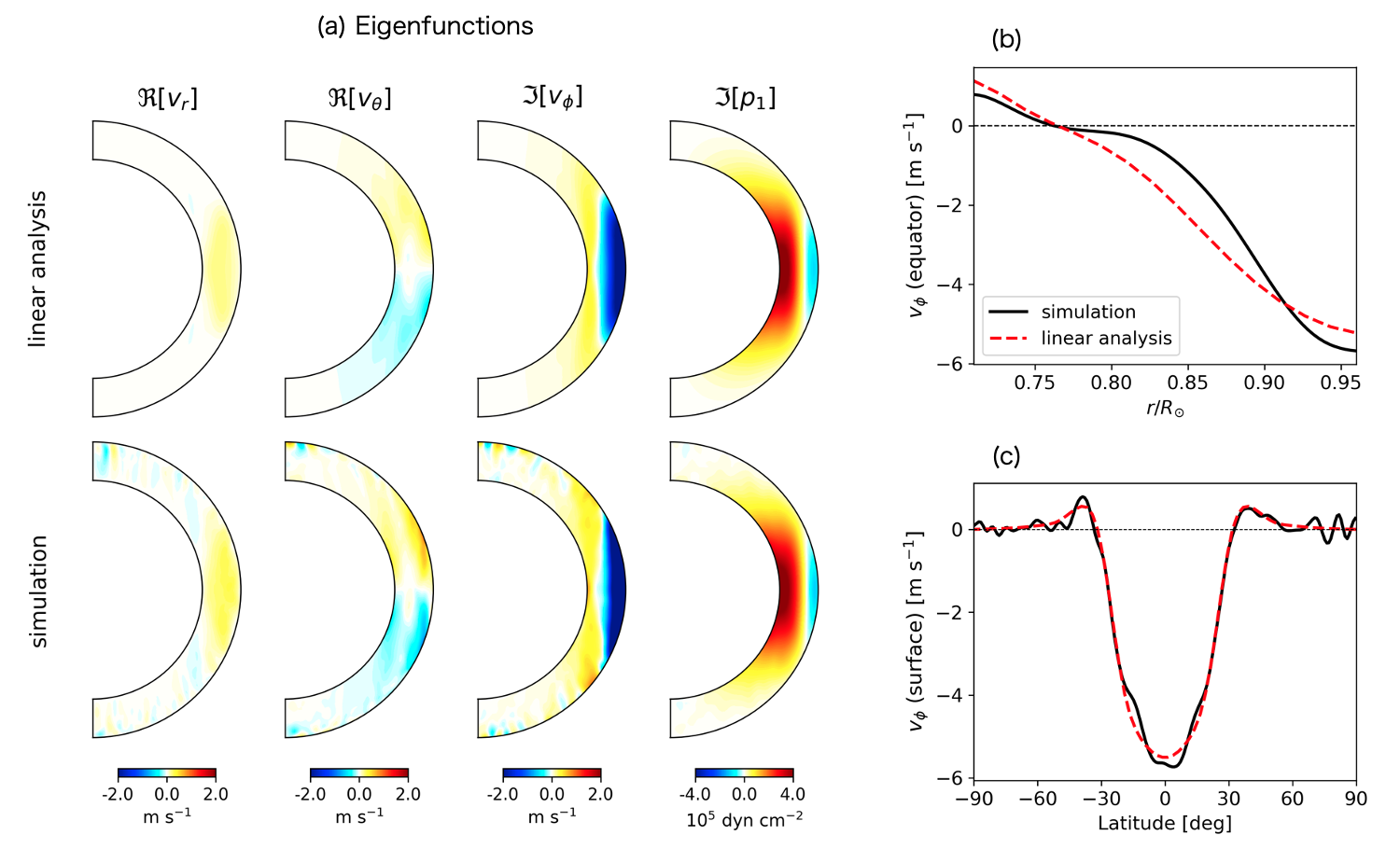}
\caption{
Plots of the spatial eigenfunctions $q_{\alpha,\mathrm{mode}}(r,\theta,m)$ for  the $\zeta_{z}$-symmetric columnar convective modes with $m=2$.
By definition, modes are of the form  $q_{\alpha,\mathrm{mode}}(r,\theta,m)\exp{\left[\ii(m\phi-\omega t)\right]}$  where $q_\alpha$ is either $v_{r}$, $v_{\theta}$, $v_{\phi}$, or $p_{1}$.
(a) Meridional cuts of the radial velocity $v_{r}$, latitudinal velocity $v_{\theta}$, longitudinal velocity $v_{\phi}$, and pressure perturbation $p_{1}$ extracted from the convection simulation (lower penals) and obtained from our linear calculation (upper panels).
(b) Radial dependence of the eigenfunction $v_{\phi}$ at the equator. 
Black solid and red dashed lines represent that of simulation and linear calculation.
(c) Latitudinal eigenfunction of $v_{\phi}$ at the surface normalized near the equator.
}
\label{fig:eigen_thRos}
\end{center}
\end{figure*}


Figure~\ref{fig:power_thRos}a shows the equatorial power spectrum ($m-\omega$ diagram) of the longitudinal velocity near the top boundary $|\tilde{v}_{{\phi},\mathrm{eq}}(0.95 R_\odot,m,\omega) |^{2}$.
The dispersion relationship for the columnar convective modes from the linear eigenmode calculation are overplotted in red.
The frequencies are given for a frame rotating with $\Omega_{0}$.
A clear power ridge can be observed in Fig.~\ref{fig:power_thRos}a, matching that from the linear analysis for $m \lesssim 10$.
The frequency of this power ridge is positive in the frame rotating with the local differential rotation rate (denoted by blue solid line), implying that the convective modes are propagating in a prograde direction \citep{miesch2008,bessolaz2011}.
Figure~\ref{fig:power_thRos}b shows the same equatorial power spectrum at fixed azimuthal order  $|\tilde{v}_{{\phi},\mathrm{eq}}(r,m=8,\omega) |^{2}$.
The strong longitudinal velocity power is localized near the surface where the compressional $\beta$-effect ($\propto H_{\rho}^{-1}$ where $H_{\rho}$ is the density scale height) is the strongest \citep[e.g.,][]{glatzmaier1981}.
At $m=8$, the mode has a linewidth of $30$ nHz and a corresponding decaying timescale of $122$ days.

We used the method described in \S~\ref{sec:svd} to extract the spatial structure of the columnar convective modes for azimuthal orders $1\le m \le 39$
\footnote{The maximum azimuthal order in our grid resolution is 192. However, we restrict our analysis to large-scale modes in a range $0\leq m \leq 39$.}.
To extract the modes, we calculated $V_{0}(\omega)$ in 
Eq.~(\ref{eqn:V0}) based on the equatorial spectrum of longitudinal velocity near surface.
Figure~\ref{fig:p1_thRos} shows the three-dimensional spatial patterns of the convective columnar modes found in our simulation for selected $m$.
For visualization purposes, the non-axisymmetric components of pressure perturbation $p_{1}$ are shown.
We note that the positive (negative) pressure perturbation $p_{1}$ is associated with negative (positive) $z$-vorticity $\zeta_{z}$ of the modes due to the strong constraint of geostrophic balance \citep[e.g.,][]{matilsky2020}.
It is clearly seen that the columnar convective modes are characterized by the north-south aligned columns across the equatorial plane.

Figure~\ref{fig:eigen_thRos} shows the extracted eigenfunctions of the convective columnar modes for $m=2$, in comparison with the results of the linear analysis.
The real eigenfunctions of $v_{r}$ and $v_{\theta}$ and the imaginary eigenfunctions of $v_{\phi}$ and $p_{1}$ are shown in Fig.~\ref{fig:eigen_thRos}a.
Figures~\ref{fig:eigen_thRos}b and c further compare the radial and latitudinal structures of the extracted mode at the equator and at the surface, respectively.
Note that the eigenfunctions extracted from the simulations are realizations of random processes, and therefore contain noise. 
The noise is visible at small spatial scales, however particularly at large scales, a great agreement can be seen between the eigenfunctions from the simulation and those of the linear analysis.
Therefore, the columnar convective modes are unambiguously identified in our simulations.
The modes can be clearly characterized by a dominant $z$-vorticity $\zeta_{z}$ that is confined outside the tangential cylinder as described in detail in \citet[][]{bekki2021_l}.
As $m$ increases, the eigenfunctions are more and more confined towards the surface and towards the equator (not shown).


\begin{figure*}[h]
\begin{center}
\includegraphics[width=0.995\linewidth]{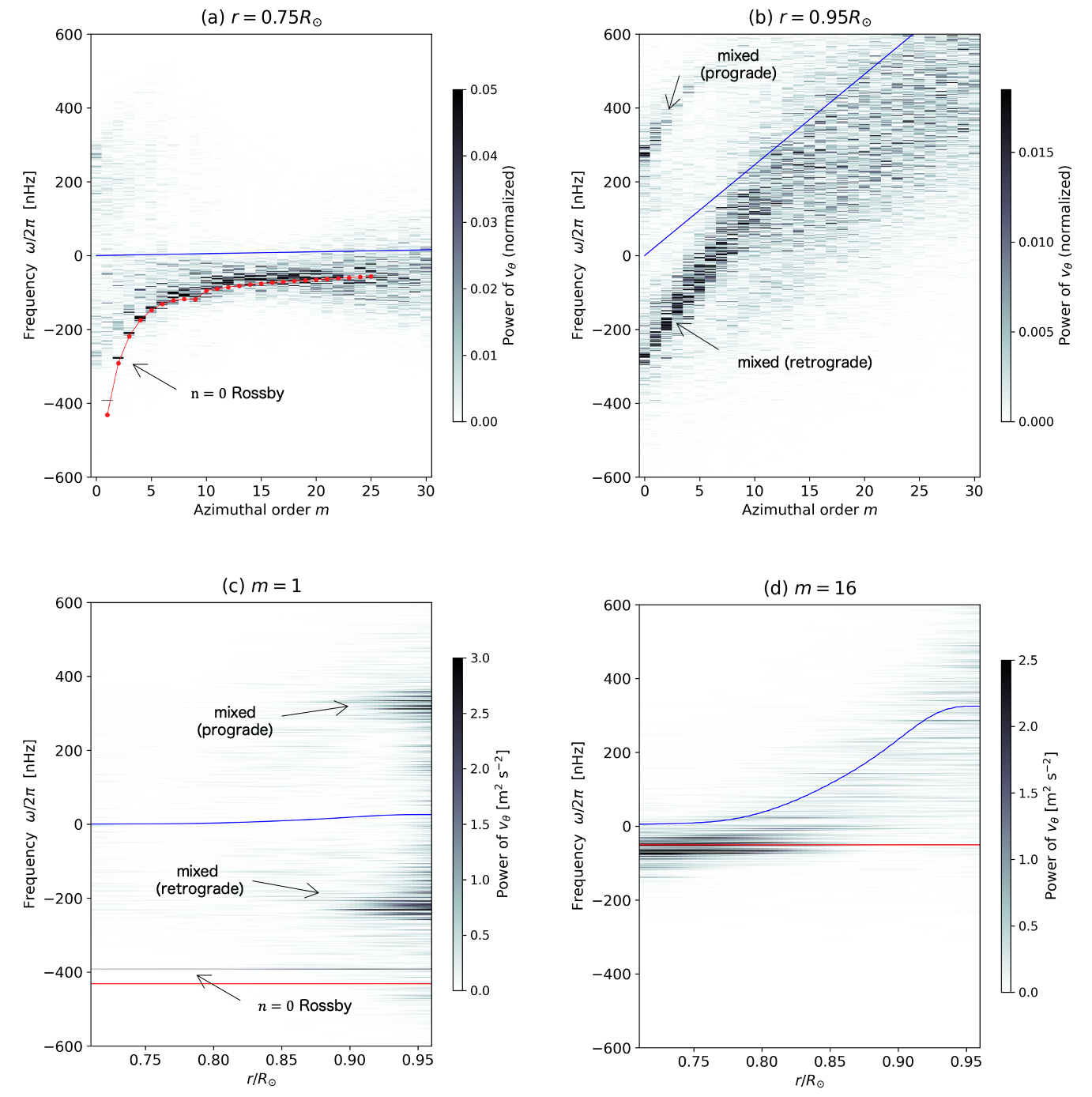}
\caption{
Power spectra of latitudinal velocity $v_{\theta}$ near the equator (averaged over $\pm 15$ deg).
Panels (a) and (b) show the $m-\omega$ diagram near the base ($r=0.75R_{\odot}$) and near the surface ($r=0.95R_{\odot}$), respectively.
The power is normalized at each $m$.
The spectra are computed in a frame rotating at $\Omega_{0}/2\pi=431.3$ nHz.
Overplotted in red line in panel (a) represents the dispersion relation of the equatorial Rossby mode with no radial node ($n=0$) obtained from the linear calculation.
The blue line represents the advection frequency of the equatorial differential rotation, $m\left[\Omega(r,\pi/2)-\Omega_{0} \right]$, at each height.
Panels (c) and (d) show the power spectra at fixed azimuthal order $m=1$, and $m=16$, respectively.
}
\label{fig:power_eqRos}
\end{center}
\end{figure*}

\begin{figure*}[h!]
\begin{center}
\includegraphics[width=0.95\linewidth]{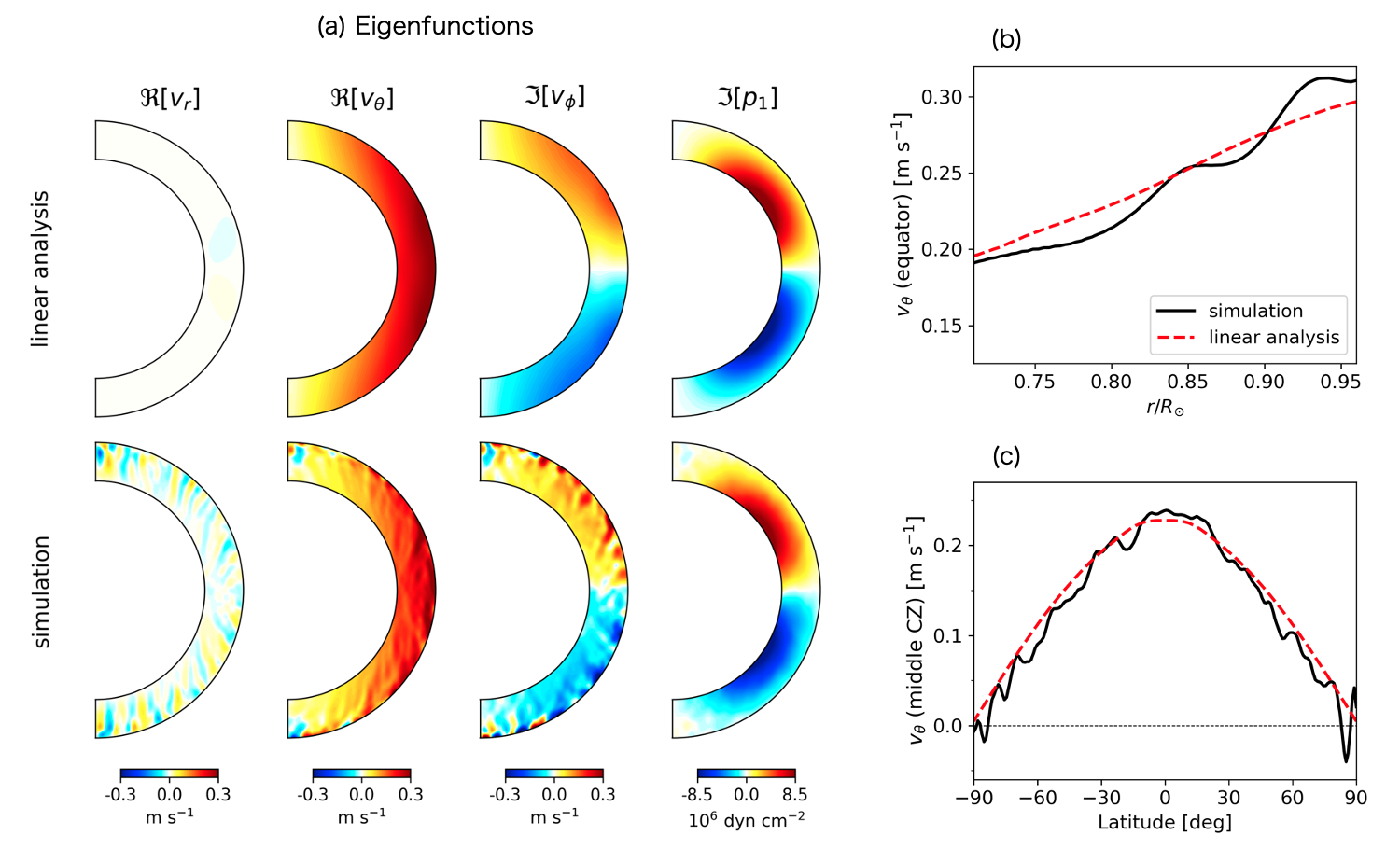}
\caption{
Eigenfunctions of the equatorial Rossby mode with no radial node ($n=0$) for $m=2$. 
(a) Meridional eigenfunctions extracted from the simulation (lower panels) and those obtained from the linear analysis (upper panels). 
The same notation is used as in Fig.~\ref{fig:eigen_thRos}a.
(b) Radial eigenfunction of $v_{\theta}$ at the equator.
Black solid and red dashed lines represent the results from simulation and linear calculation, respectively.
(c) Latitudinal eigenfunction of $v_{\theta}$ in the middle convection zone.
}
\label{fig:eigen_n0eqRos}
\end{center}
\end{figure*}


\subsection{Equatorial Rossby modes} \label{sec:eqRos}

In this section, we present the same modal analysis for the equatorial Rossby modes (r modes) where radial vorticity $\zeta_{r}$ is symmetric across the equator.
In this study, we use the latitudinal velocity $v_{\theta}$ near the equator for our spectral analysis which is a good representative of the equatorial Rossby modes.
Figures~\ref{fig:power_eqRos}a and b show the equatorial power spectra ($m-\omega$ diagram) of latitudinal velocity near the base $|\tilde{v}_{\theta,\mathrm{eq}}(0.715R_{\odot},m,\omega)|^{2}$ and near the surface $|\tilde{v}_{\theta,\mathrm{eq}}(0.95R_{\odot},m,\omega)|^{2}$, respectively. 
The dispersion relation of the equatorial Rossby modes with no radial node ($n=0$) obtained from our linear analysis are shown in red points in Fig.~\ref{fig:power_eqRos}a.
A clear power ridge can be seen along the linear dispersion relation near the base, whereas two distinct ridges are found in the surface power spectrum (denoted as ``mixed prograde'' and ``mixed retrograde'' in Fig.\ref{fig:power_eqRos}b).

Figures~\ref{fig:power_eqRos}c and d show the same equatorial power spectra of $v_{\theta}$ at fixed azimuthal orders $|\tilde{v}_{\theta,\mathrm eq}(r,m=1,\omega)|^{2}$ and $|\tilde{v}_{\theta,\mathrm eq}(r,m=16,\omega)|^{2}$, respectively.
As shown in Fig.~\ref{fig:power_eqRos}c, three distinct modes exist at low-$m$ regime; 
\begin{itemize}
\item[$\bullet$] A retrograde-propagating mode that exists globally in radius at $\omega/2\pi\approx-395$ nHz.
Its mode frequency is close to the theoretical eigenfrequency of the $n=0$ Rossby mode predicted in linear analysis (denoted by red line) and is independent of height.
This mode is undoubtedly identified as the $n=0$ equatorial Rossby mode and will be discussed in \S\ref{sec:eqRos-lowm}.
\item[$\bullet$] A retrograde-propagating mode localized near the surface at $\omega/2\pi\approx-230$ nHz is also seen.
This mode is identified as the equatorial Rossby mode with one radial node ($n=1$) and will be discussed in \S\ref{sec:mixed}.
\item[$\bullet$] A prograde-propagating mode localized near the surface at $\omega/2\pi\approx320$ nHz is also apparent.
This mode is identified as the north-south $\zeta_{z}$-antisymmetric columnar convective mode and will also be discussed in \S\ref{sec:mixed}.
\end{itemize}
At higher $m$, on the other hand, most of the power is concentrated near the bottom convection zone at frequencies close to those of the $n=0$ Rossby modes from the linear analysis (denoted by red line) as shown in Fig.~\ref{fig:power_eqRos}d.
Properties of these high-$m$ Rossby modes will be discussed in \S\ref{sec:eqRos-highm}.


\subsubsection{Rossby modes with $n=0$ and $m\leq 4$} \label{sec:eqRos-lowm}

At low $m$, equatorial Rossby modes can be unambiguously found in our simulation.
Figure~\ref{fig:eigen_n0eqRos}a shows the extracted eigenfunctions of $m=2$ mode as an example. 
The associated flow motion is mostly $r$-vortical and quasi-toroidal, i.e., $v_{r}\approx0$.
Geostrophical balance is well established by positive (negative) $p_{1}$  in a region where $\zeta_{r}$ is negative (positive) in the northern (southern) hemisphere.
The radial component of the Coriolis force is balanced by the radial pressure gradient force. 
For $m \le 4$, almost all the power is in real part of the eigenfunction of $v_{\theta}$ ($\Re[v_{\theta}]$) and the imaginary part for $v_{\phi}$ ($\Im[v_{\phi}]$):
The other components of the eigenfunctions (for example $\Im[v_{\theta}]$ or $\Re[v_{\phi}]$) are small and consistent with noise.
For comparison, we also show the eigenfunctions of $m=2$ equatorial Rossby mode with no radial node ($n=0$) obtained from our linear calculation in the uppper panel of Fig.~\ref{fig:eigen_n0eqRos}a.
A very good agreement can be seen between the extracted eigenfunctions from the nonlinear simulation and those of the linear calculation.
Figures~\ref{fig:eigen_n0eqRos}b and c further compare the radial and latitudinal structures of the eigenfunction of $v_{\theta}$.
The radial eigenfunction roughly shows a monotonic increase towards the surface as expected from the analytical solution for the ideal case of inviscid and uniformly-rotating sphere, $v_{\theta} \propto r^{m}$ \citep[e.g.,][]{saio1982}.
Similarly, the latitudinal eigenfunction also roughly follows the analytical solution for the ideal case, $v_{\theta} \propto \sin^{m-1}{\theta}$.

It is noteworthy that the $m=1$ equatorial Rossby mode shown in Fig.~\ref{fig:power_eqRos}c has a flow vorticity that is uniform in the frame of the mode and points in a direction perpendicular to the solar rotation axis.
This particular mode is often called the ``spin-over inertial mode'' \citep[e.g.,][]{greenspan1968} and has been extensively studied in the context of planetary cores \citep[e.g.,][]{triana2012,rekier2022}.

We also note that these equatorial Rossby modes at low $m$ are very long-lived.
For instance, the $m=4$ mode has a linewidth of $7.5$ nHz and the corresponding lifetime of about $500$ days.
The modes with $m<4$ have much longer lifetimes as their linewidths are too small to be well resolved and to be fitted with Lorentzian.

\begin{figure*}[]
\begin{center}
\includegraphics[width=0.95\linewidth]{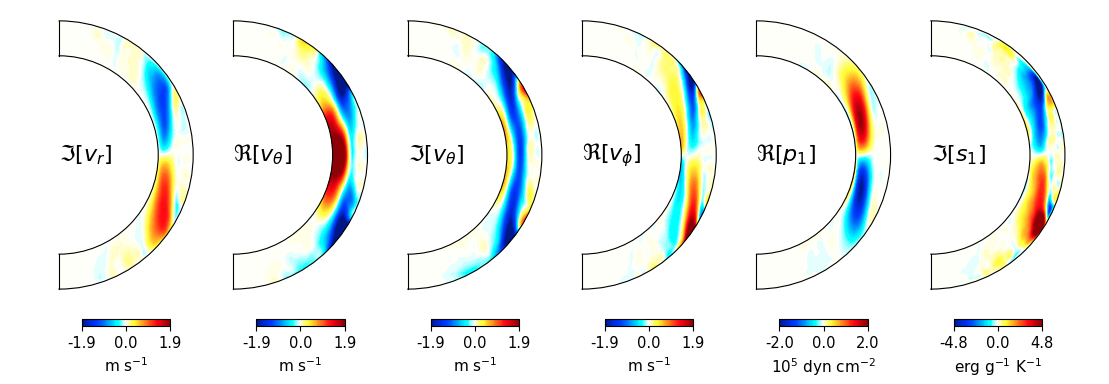}
\caption{
Extracted eigenfunctions of the $n=0$ equatorial Rossby modes at $m=24$.
Eigenfunctions of radial velocity (imaginary) $\Im[v_{r}]$, latitudinal velocity (both real and imaginary) $\Re[v_{\theta}]$, $\Im[v_{\theta}]$, longitudinal velocity (real) $\Re[v_{\phi}]$, pressure perturbation (real) $\Re[p_{1}]$, and entropy perturbation (imaginary) $\Im[s_{1}]$ are shown from left to right.
The real and imaginary phases are determined in a way that $\Re[v_{\theta}]$ at the base takes its maximum at the equator.
}
\label{fig:eigen_eqRosall}
\end{center}
\end{figure*}

\begin{figure}[]
\begin{center}
\includegraphics[width=\linewidth]{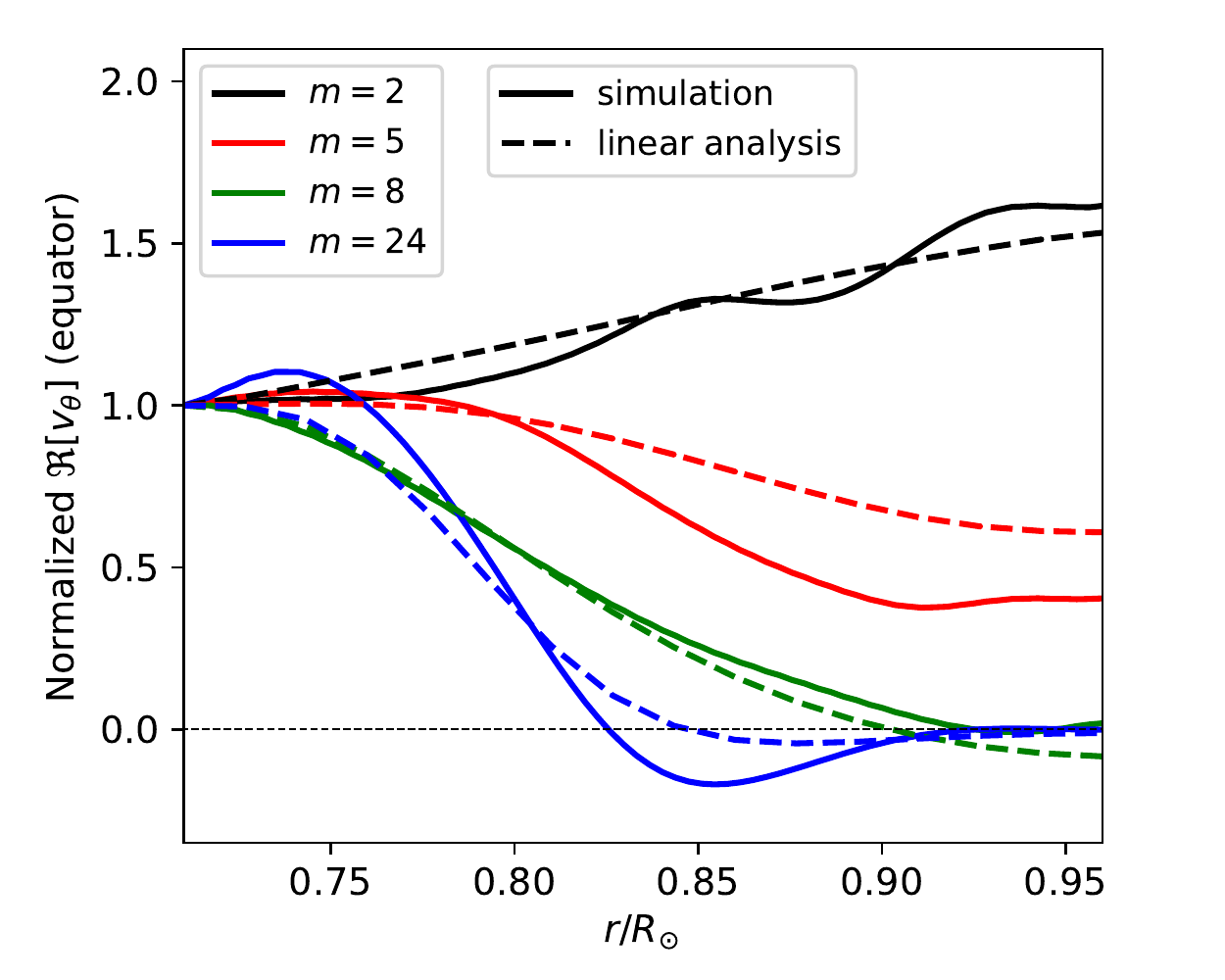}
\caption{Radial structure of the real eigenfunctions of latitudinal velocity $\Re[v_{\theta}]$ at the equator.
The eigenfunctions are normalized to unity at the base.
Different colors represent different azimuthal orders $m$.
Solid and dashed lines denote those extracted from the nonlinear simulation and from the linear eigenmode calculation, respectively.
}
\label{fig:vq_eqRos}
\end{center}
\end{figure}
\begin{figure}[]
\begin{center}
\includegraphics[width=\linewidth]{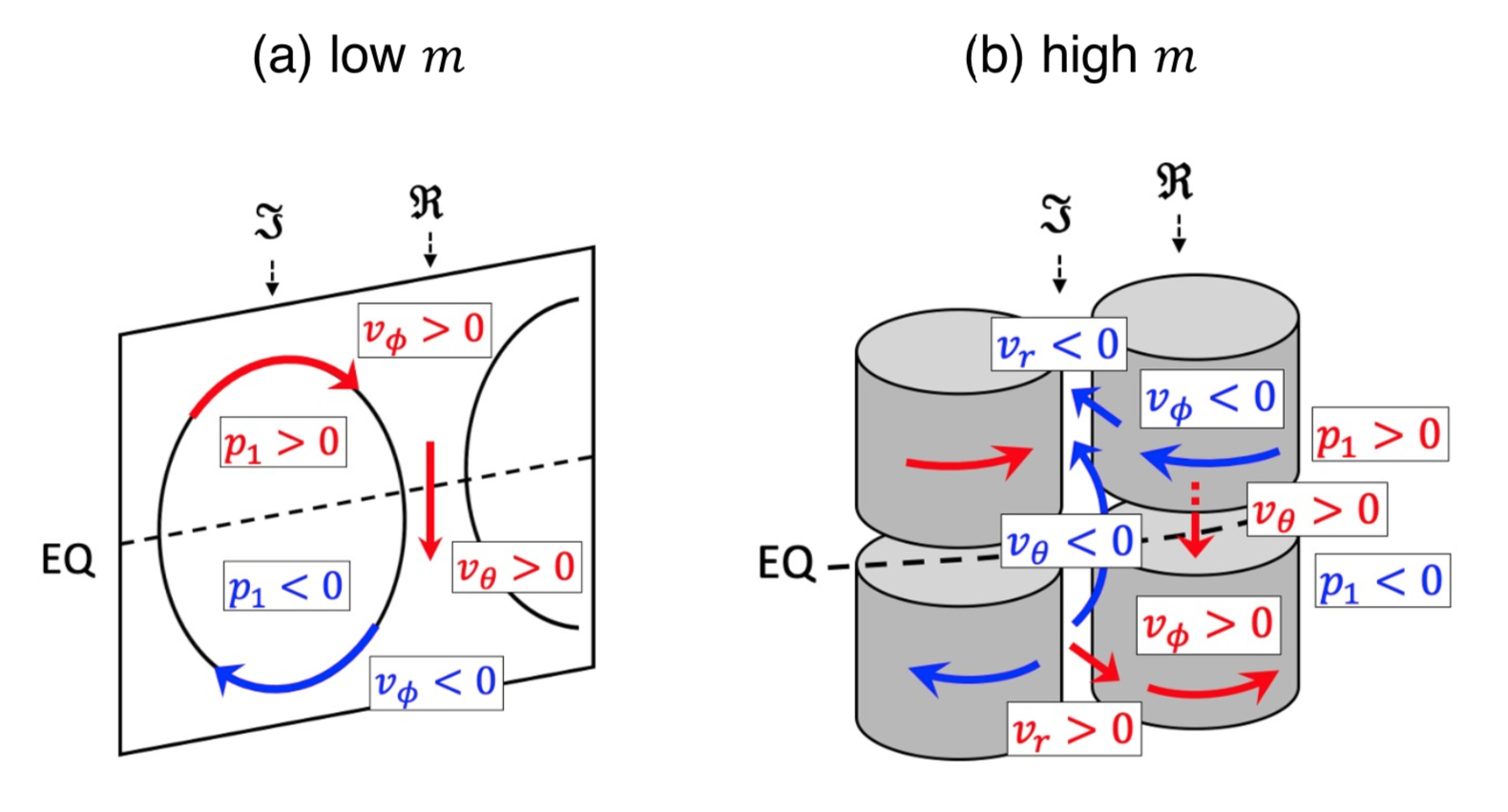}
\caption{
Schematic illustration of the flow structure of the $n=0$ equatorial Rossby modes at (a) low-$m$ and (b) high-$m$ regimes.
}
\label{fig:illust_eqRos}
\end{center}
\end{figure}

\begin{figure*}[]
\begin{center}
\includegraphics[width=0.98\linewidth]{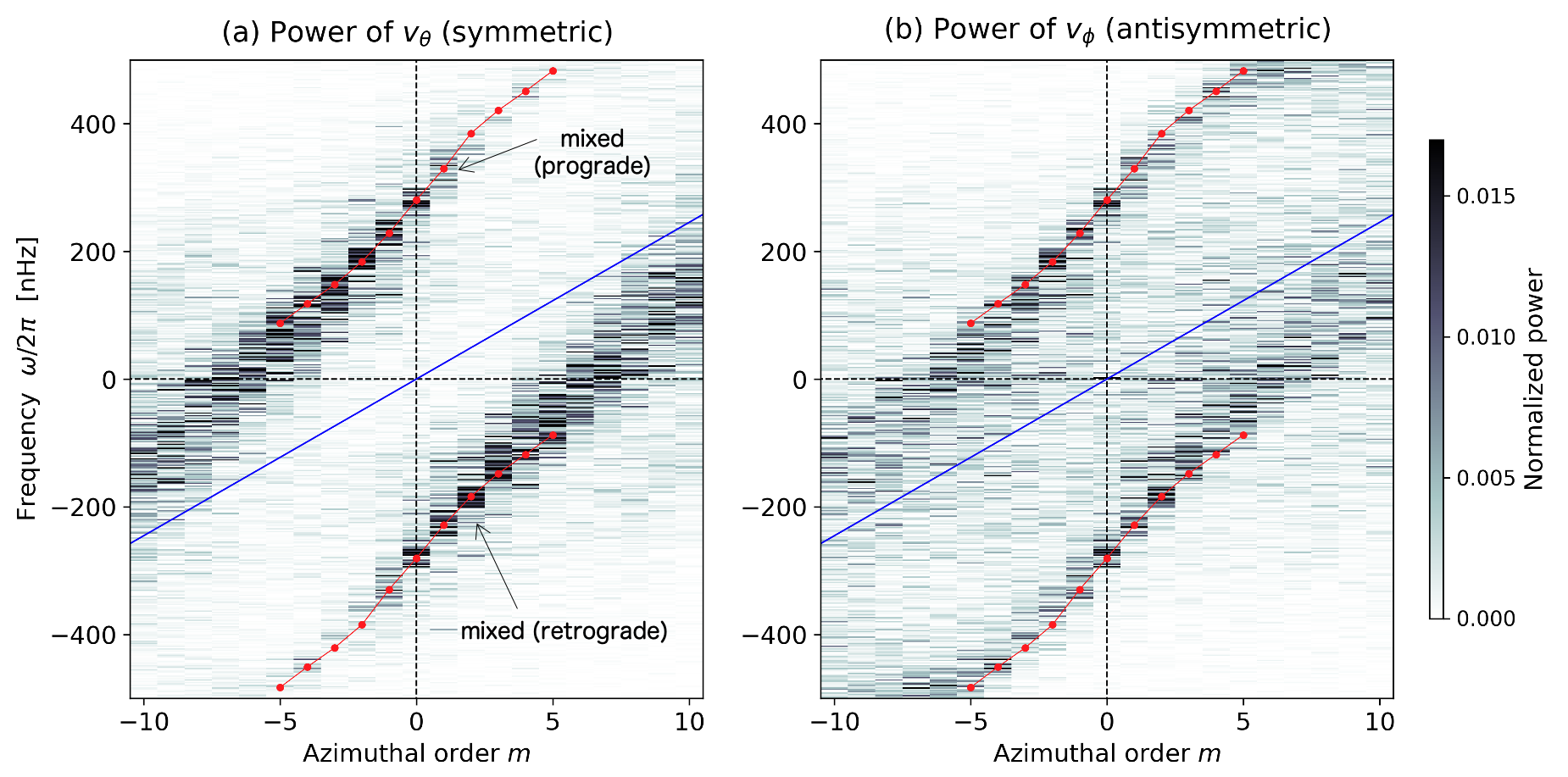}
\caption{
Power spectra of horizontal velocities near the surface extended to negative azimuthal orders ($m<0$). 
(a) Equatorial power spectrum of north-south symmetric component of $v_{\theta}$ near the top boundary $r=0.95R_{\odot}$, which is the same as Fig.~\ref{fig:power_eqRos}b. 
The power is normalized at each $m$.
Shown in red points are the frequencies of the ``mixed Rossby modes'' obtained from our linear analysis. 
The blue line represents the advection frequency of the equatorial differential rotation, $m\left[\Omega(0.95R_{\odot},\pi/2)-\Omega_{0} \right]$.
(b) Equatorial power spectrum of north-south antisymmetric component of $v_{\phi}$ near the top boundary $r/R_{\odot}=0.95$.
 }
\label{fig:power_mixed}
\end{center}
\end{figure*}

\begin{figure*}[]
\begin{center}
\includegraphics[width=0.92\linewidth]{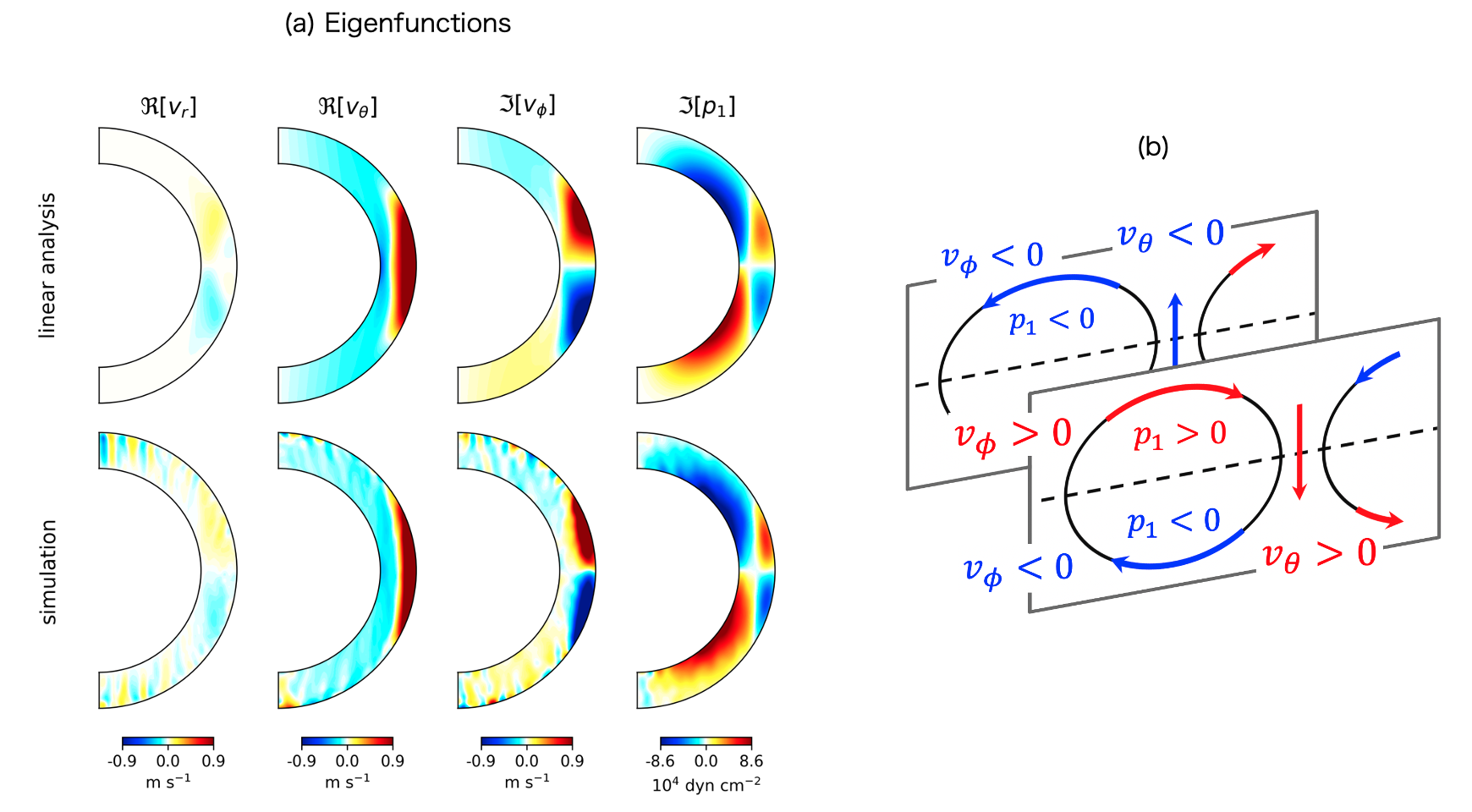}
\caption{
(a) Eigenfunctions of the retrograde-propagating mixed Rossby mode ($n=1$ equatorial Rossby mode) at $m=2$. 
Lower and upper panels show the results extracted from the simulation and those obtained from the linear analysis. 
(b) Schematic illustration of this mode.
}
\label{fig:eigen_n1eqRos}
\end{center}
\end{figure*}

\begin{figure*}[]
\begin{center}
\includegraphics[width=0.92\linewidth]{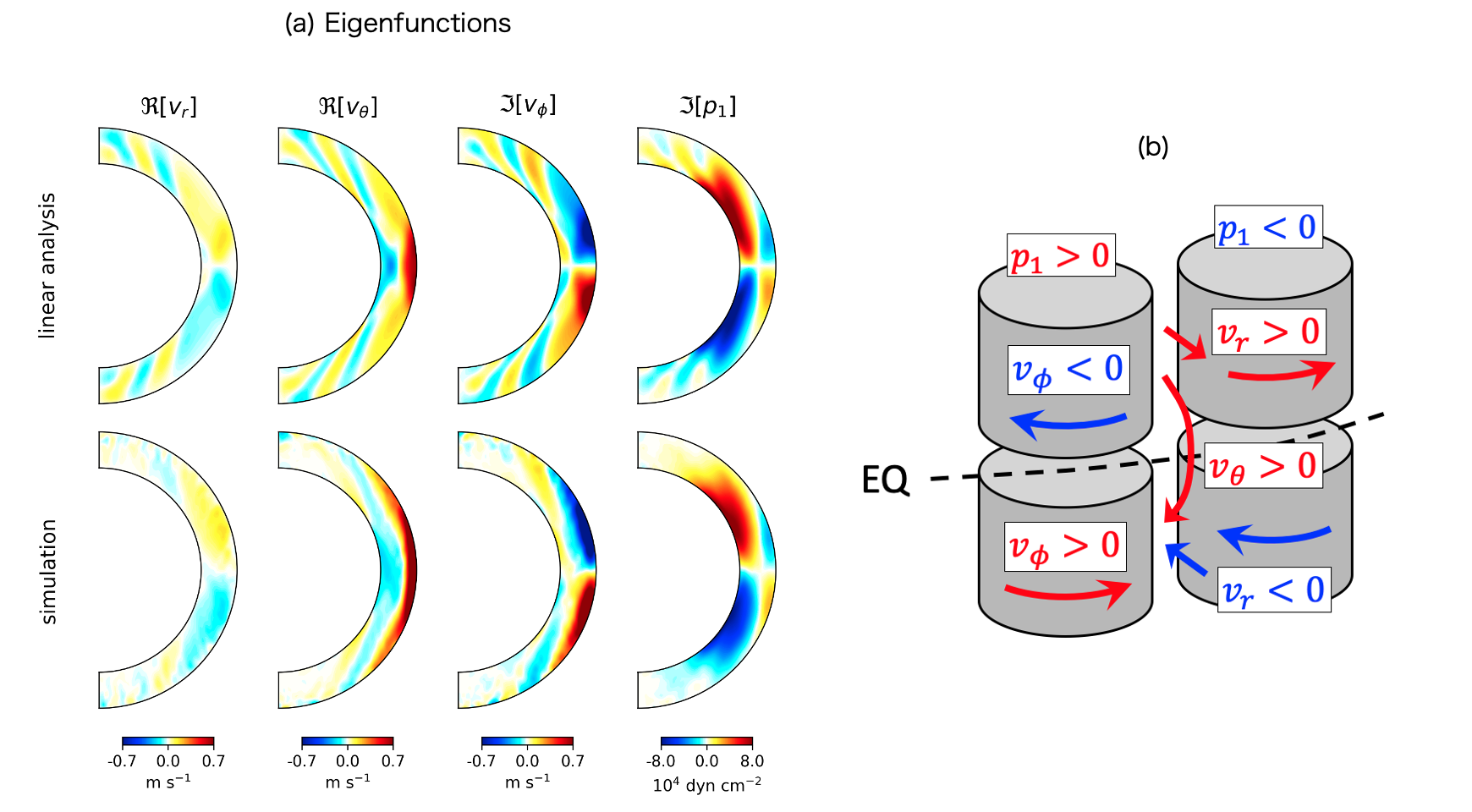}
\caption{
(a) Eigenfunctions of the prograde-propagating mixed Rossby mode (north-south $\zeta_{z}$-antisymmetric columnar convective mode) at $m=2$. 
Lower and upper panels show the results extracted from the simulation and those obtained from the linear analysis. 
(b) Schematic illustration of this mode.
}
\label{fig:eigen_asthRos}
\end{center}
\end{figure*}


\subsubsection{Rossby modes with $n=0$ and $m>4$} \label{sec:eqRos-highm}

As $m$ increases, the eigenfunctions of the $n=0$ equatorial Rossby modes significantly deviate from the well-known analytical expression of, $v_{\theta} \propto r^{m} \sin^{m-1}{\theta}$.
Figure~\ref{fig:eigen_eqRosall} shows the eigenfunctions at $m=24$ that is extracted from the convection simulation along the power ridge shown in Fig.~\ref{fig:power_eqRos}a.
We emphasize that the eigenfunctions become essentially complex, with the complex phase for each variable being a function of $r$ and $\theta$.
For large azimuthal order $m$, the modes cease to be quasi-toroidal, i.e., the radial motions become substantial and the modes are partially convectively driven.
This can be confirmed by the fact that, in Fig.~\ref{fig:eigen_eqRosall}, $v_{r}$ and $s_{1}$ have the same sign at the same phase in both hemispheres.
The transport properties of thermal energy will later be discussed in detail in \S~\ref{sec:transport_Fe}.

It is also found that the $n=0$ Rossby modes are more and more confined towards the base of the convection zone as $m$ increases.
This is clearly illustrated in Fig.~\ref{fig:vq_eqRos} where the normalized eigenfunctions of $\Re[v_{\theta}]$ at the equator are plotted against radius for different $m$.
For $m \geq 4$, the radial eigenfunctions become decreasing functions in radius, which clearly disagrees with the analytical solution of Rossby modes in the ideal case (uniform rotation and no viscosity) \citep[][]{saio1982}.
In the ideal case, $r^{m}$ dependence is required in order to balance the radial component of the Coriolis force by the radial pressure gradient.
We attribute this discrepancy to the strong viscous diffusion in our simulation:
When strong diffusivity is present, the radial force balance is drastically changed.
This has already been pointed out by \citet[][]{bekki2021_l} who showed, in the linear regime, that the eigenfunctions of $n=0$ Rossby modes tend to be localized near the base of the convection zone when the turbulent viscosity is above $\approx 10^{12}$ cm$^{2}$~s$^{-1}$.
To confirm this, we show in Fig.~\ref{fig:vq_eqRos} the results from our linear calculations by dashed lines that can well explain this trend.
In fact, the effective diffusivity in our nonlinear simulation is  substantially enhanced by the turbulent (eddy) momentum mixing by the convective flows.
In Appendix \ref{appendix:B}, we estimate the turbulent diffusivity due to the small-scale convective motions in the simulation to be $\approx 3\times 10^{12}$ cm$^{2}$ s$^{-1}$ throughout the convection zone, which is larger than the explicit viscosity value of $10^{12}$ cm$^{2}$ s$^{-1}$. 
Therefore, the effective viscosity is dominated by the eddy diffusion due to the stochastic convective motions on resolved scales.
For more detailed discussions on the effect of turbulent diffusion on Rossby modes, see \S~5 in \citet{bekki2021_l}.

Figure~\ref{fig:eigen_eqRosall} also reveals that, at high $m$, the flow motions are characterized by $z$-vortices along with the tangential cylinder, as manifested by the eigenfunctions of $\Im[v_{r}]$, $\Im[v_{\theta}]$, and $\Re[v_{\phi}]$.
The essential difference from the columnar convective modes discussed in \S\ref{sec:thRos} is here that $\zeta_{z}$ is north-south antisymmetric across the equator.
This is schematically illustrated in Fig.~\ref{fig:illust_eqRos}.
These modes should also be distinguished from the prograde-propagating ``mixed Rossby modes'' (that we will discuss in \S~\ref{sec:mixed}) where the power is strongly localized near the surface and not at the tangential cylinder.

Note that these high-$m$ equatorial Rossby modes have generally broader linewidths (thus shorter lifetimes) than those with low $m$.
For instance, the mode with $m=15$ has a linewidth of about $25$ nHz.
This is in fact comparable to those observed in the Sun \citep[the linewidth of the $m=15$ mode observed on the Sun is reported to be $\approx 10$--$40$ nHz, see][]{loeptien2018,liang2019,proxauf2020}.

\subsection{Mixed Rossby modes} \label{sec:mixed}

Figure~\ref{fig:power_mixed}a shows the same equatorial power spectrum near the surface as Fig.~\ref{fig:power_eqRos}b except that we extend the spectrum to negative azimuthal orders ($m<0$) considering the symmetry with respect to the origin; $|\tilde{v}_{\theta,\mathrm{eq}}(-m,-\omega)|^{2}=|\tilde{v}_{\theta,\mathrm{eq}}(m,\omega)|^{2}$.
We also show the equatorial power spectrum of the north-south antisymmetric component of $v_{\phi}$ in Fig.~\ref{fig:power_mixed}b.
Note that both Fig.~\ref{fig:power_mixed}a and b have the same symmetry and thus represent the same modes.
This enables us to better see the connection between the two distinct power ridges that are each denoted by ``mixed (retrograde)'' and ``mixed (prograde)'':
These two oppositely-propagating modes form a single continuous power ridge across $m=0$, implying that they are essentially mixed with each other.
The axisymmetric mode ($m=0$) is considered to be an inertial mode trapped inside the spherical shell \citep[e.g.,][]{rieutord2001,rieutord2018} and has an oscillation frequency of about $\omega/2\pi\approx\pm 280$ nHz.

Figure~\ref{fig:eigen_n1eqRos}a shows the extracted eigenfunctions of the retrograde-propagating mode at $m=2$.
The retrograde-propagating mode can be classified as an equatorial Rossby mode with one radial node $n=1$ in the middle convection zone as depicted in Fig.~\ref{fig:eigen_n1eqRos}b.
We note that the nodal plane is more cylindrical than radial outside the tangential cylinder.
The associated motion is dominantly $r$-vortical near the surface, but unlike the $n=0$ equatorial Rossby modes, non-negligible radial velocities are involved.
Similarly, the extracted eigenfunctions of the prograde-propagating mode at $m=2$ are shown in Figure~\ref{fig:eigen_asthRos}a.
The prograde-propagating mode can be classified as a north-south $\zeta_{z}$-antisymmetric columnar convective mode as illustrated in Fig.~\ref{fig:eigen_asthRos}b.
Given that both of these modes follow the same dispersion relationship, we call them ``mixed Rossby modes'' in this paper.

The existence of the ``mixed Rossby modes'' was first pointed out in \citet[][]{bekki2021_l}.
Their dispersion relation asymptotically approaches to that of $n=0$ equatorial Rossby mode for large $m$ with negative $\omega$ (retrograde modes) and to that of the north-south $\zeta_{z}$-symmetric columnar convective mode for large $m$ with positive $\omega$ (prograde modes).
The coupling between these two oppositely-propagating modes can be understood by analogy to the well-known mixed Rossby-gravity waves (sometimes called Yanai waves) in the geophysical context \citep[e.g.,][]{matsuno1966,vallis2006}
whose dispersion relation is asymptotic to that of classical Rossby waves for large $m$ with negative $\omega$ (retrograde) and asymptotic to that of inertia-gravity waves for large $m$ with positive $\omega$ (prograde).

To further support the identification of these mixed Rossby modes, we compute the dispersion relation and the corresponding eigenfunctions using the linear eigenvalue solver.
The computed dispersion relation is shown by red points in Fig.~\ref{fig:power_mixed} for $m \leq 5$, which nicely agrees with the power ridge in the simulated spectrum.
We also show in Fig.~\ref{fig:eigen_n1eqRos}a and Fig.~\ref{fig:eigen_asthRos}a the eigenfunctions of the $n=1$ equatorial Rossby mode and the north-south $\zeta_{z}$-antisymmetric columnar convective mode at $m=2$ obtained from the linear calculation.
The agreement between the simulation and linear theory is striking.

\begin{figure*}[h]
\begin{center}
\includegraphics[width=0.99\linewidth]{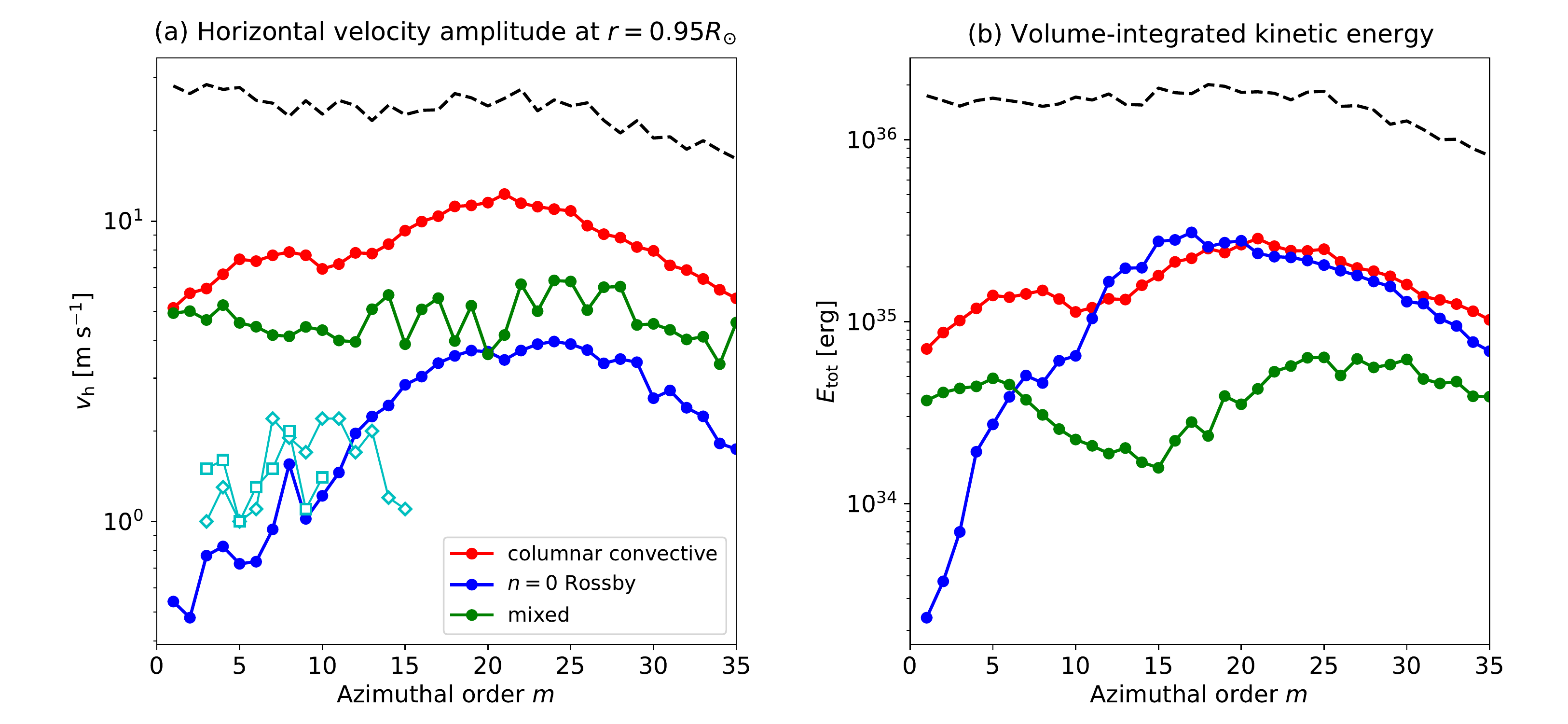}
\caption{
(a) Maximum horizontal velocity $v_{\mathrm{h}}$ of the equatorial modes near the top boundary $r=0.95R_{\odot}$ at each azimuthal order $m$.
Red, blue, and green points represent the $\zeta_{z}$-symmetric columnar convective modes, $n=0$ equatorial Rossby modes, and the mixed Rossby modes, respectively.
Black dashed line represents the overall power of the convection simulation (including modes at high latitudes and stochastic convective motions).
Cyan diamonds and squares denote the (rms) horizontal velocity amplitudes of the observed Rossby modes near the solar surface obtained by \citet{liang2019} and \citet{gizon2021}, respectively.
(b) Spectra of volume-integrated kinetic energy of the equatorial modes.
}
\label{fig:amplitude}
\end{center}
\end{figure*}

\section{Transport properties of low-frequency modes} \label{sec:transport}

\begin{table*}[]
  \begin{center} 
\caption{Amplitudes and lifetimes of the equatorial vorticity modes in our simulation} \label{table:summary}
\small
\begin{tabular}{cccccccccccc} 
\toprule
\toprule
\multirow{2}{*}{Classification}  & \ \ \ & kin. energy & \multicolumn{2}{c}{$v_{\mathrm{max}}$ [m s$^{-1}$]} & &
\multicolumn{2}{c}{$\tau_{\mathrm{mode}}$ [days]} \\ 
      \cline{4-5} \cline{7-8}
& \ \ \  &  $E_{\mathrm{mode}}/E_{\mathrm{all}}$ [$\%$] &  $m=2$ & $m=16$ & 
& $m=2$ & $m=16$ 
 \\
 \midrule
 \midrule
Columnar convective ($\zeta_{z}$-sym) & &   7.12 & 5.75 & 9.96 && 78.4 & 30.4  \\
Equatorial Rossby ($n=0$) & &    6.07 & 0.48 & 3.68 && $> 903$ & 164.7 \\
 \multirow{2}{*}{$\left.\begin{array}{c} \mathrm{Columnar \ convective} \ (\zeta_{z}\mathchar`-\mathrm{antisym})  \\
 \mathrm{Equatorial \ Rossby} \ (n=1)  \end{array}\right\rbrace \mathrm{mixed} $} && 1.18 & 1.99 & 1.31 && 56.9 & 20.6  \\
 & &   0.51 & 3.03 & 0.92 && 90.2 & 11.2 \\
\bottomrule
\end{tabular}
\end{center}
\vspace{-1.0\baselineskip}
\tablefoot{
Each row refers to a set of modes with different $m$ values.
Shown in 2nd column are the volume-integrated kinetic energies of each mode integrated over $m$ ($E_{\mathrm{mode}}$) normalized by the total fluctuating kinetic energy $E_{\mathrm{all}}$.
Shown in 3rd and 4th columns are the maximum flow amplitudes $v_{\mathrm{max}}$ of the extracted modes at $m=2$ and $m=16$, respectively.
Shown in 5th and 6th columns are the $e$-folding lifetimes $\tau_{\mathrm{mode}}=2/\Gamma_{\mathrm{mode}}$ of the $m=2$ and $m=16$ modes, where $\Gamma_{\mathrm{mode}}/2\pi$ is the mode linewidth (full width at half maximum).
} 
 \end{table*}

\subsection{Mode amplitudes} \label{sec:amplitude}

So far, we have investigated the eigenfunctions and eigenfrequencies of the north-south $\zeta_{z}$-symmetric columnar convective modes, the $n=0$ equatorial Rossby modes, and the ``mixed Rossby modes''.
In this section, we discuss how significant these modes are for transport processes in our nonlinear simulation. 

Figure~\ref{fig:amplitude}a shows the spectra of the maximum horizontal velocity $v_{\mathrm{h}}=\sqrt{v_{\theta}^{2}+v_{\phi}^{2}}$ of the equatorial modes near the surface.
It is clearly seen that the north-south $\zeta_{z}$-symmetric columnar convective modes are the most dominant in power:
They account for about $10-30~\%$ of the total velocity power of the simulation near the surface (black dashed line).
When observed at the surface, the $n=0$ equatorial Rossby modes are much weaker than the columnar convective modes and the ``mixed Rossby modes''.
Nonetheless, their amplitudes are comparable to those observed on the Sun \citep[][]{liang2019,gizon2021}
This may imply that these equatorial Rossby modes are both excited and damped by the turbulent convective motions.

Figure~\ref{fig:amplitude}b shows, on the other hand, the spectra of the volume-integrated kinetic energies of these modes.
It is shown that, despite the weak amplitudes in the surface spectrum, the total kinetic energy of the $n=0$ equatorial Rossby modes become significant:
It is much larger than that of the mixed modes for $m>6$, and become comparable to that of the $\zeta_{z}$-symmetric columnar convective modes for $m \geq 10$.
This reflects the fact the $n=0$ Rossby modes are concentrated near the base of the convection zone at high $m$ where the background density is substantial.
As shown in the 2nd column of Table~\ref{table:summary}, the $\zeta_{z}$-symmetric columnar convective modes and the $n=0$ Rossby modes have about $7\%$ and $6\%$ of the total fluctuating ($m \ne 0$) kinetic energy of the simulation, respectively.

\subsection{Thermal energy transport} \label{sec:transport_Fe}

\begin{figure*}[]
\begin{center}
\includegraphics[width=0.95\linewidth]{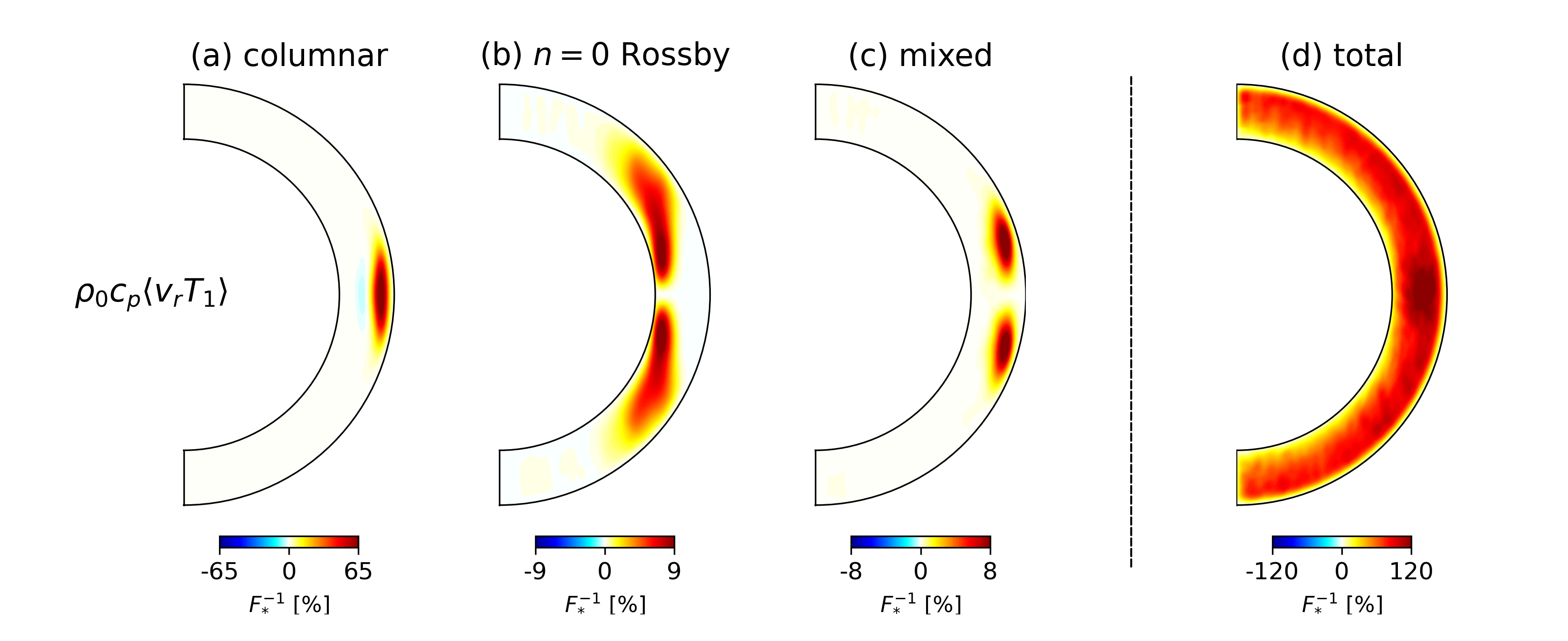}
\caption{
Enthalpy fluxes $F_{\mathrm{e}}$ associated with the extracted modes in our simulations summed over $m=1-39$.
Panels (a), (b), and (c) show those of the $\zeta_{z}$-symmetric columnar convective modes, the $n=0$ equatorial Rossby modes, and the mixed Rossby modes, respectively.
In panel (d), the total enthalpy flux (including other modes and small-scale convection) is shown.
The fluxes are normalized by the injected energy flux $F_{*}=L_{*}/(4\pi r^{2})$.
}
\label{fig:Fe}
\end{center}
\end{figure*}
\begin{figure*}[]
\begin{center}
\includegraphics[width=0.95\linewidth]{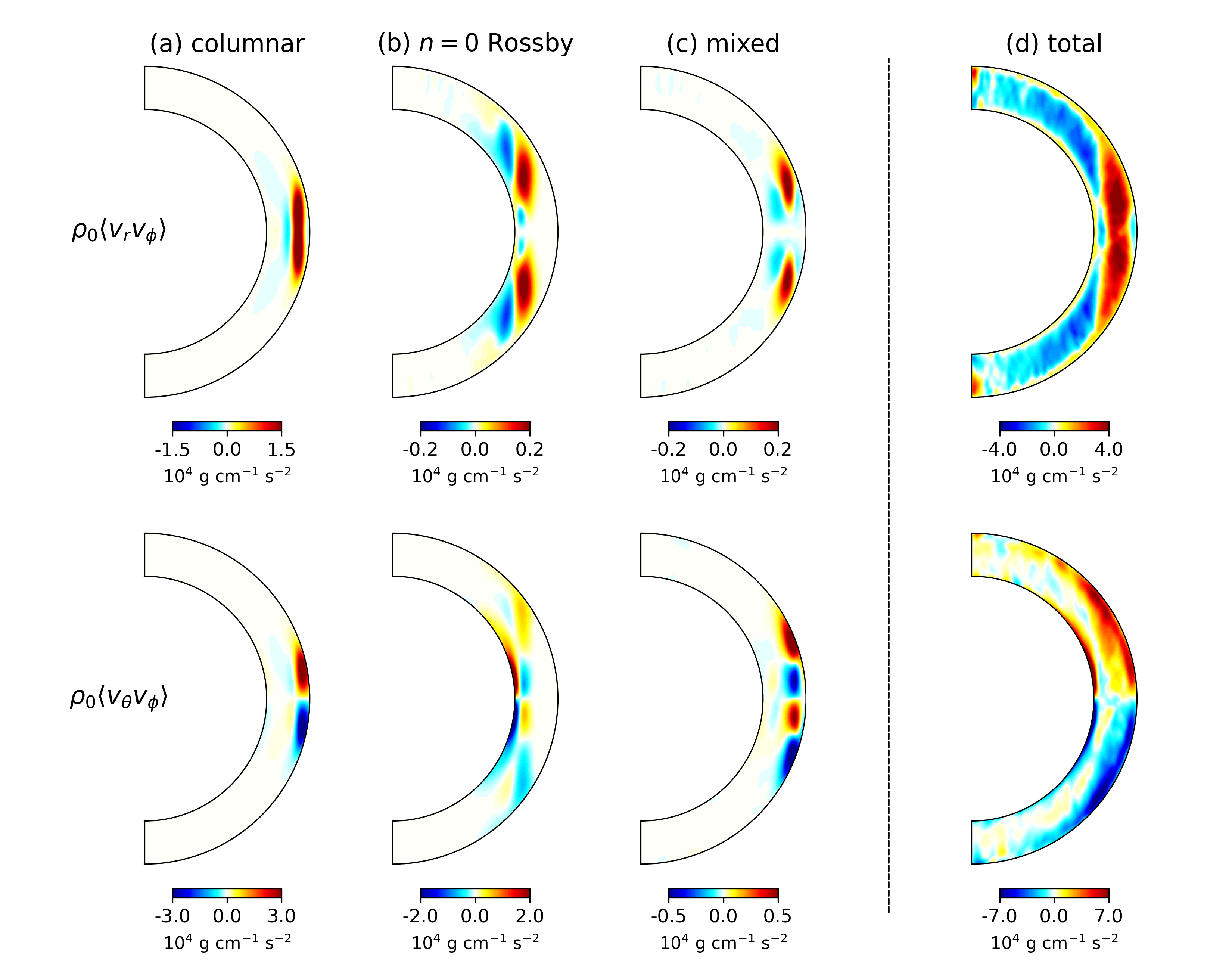}
\caption{
Reynolds stresses associated with the extracted modes in our simulations summed over $m=1-39$.
Panels (a), (b), and (c) show those of the $\zeta_{z}$-symmetric columnar convective modes, the $n=0$ equatorial Rossby modes, and the mixed Rossby modes, respectively.
In panel (d), the total Reynolds stresses (including other modes and small-scale convection) is shown.
Upper and lower panels correspond to $\rho_{0}\langle v_{r}v_{\phi}\rangle$ and $\rho_{0}\langle v_{\theta}v_{\phi}\rangle$.
}
\label{fig:RS}
\end{center}
\end{figure*}

To examine the properties of the thermal energy transport, we compute the (radial) enthalpy flux $F_{e}$ for each extracted mode as
\begin{eqnarray}
&& F_{\mathrm{e}}=\rho_{0}c_{\mathrm{p}}\langle v_{r} T_{1} \rangle,
\end{eqnarray}
where $\langle \rangle$ denotes the longitudinal average and $T_{1}$ is temperature perturbation
\begin{eqnarray}
&& T_{1}=\left[ \frac{\gamma-1}{\gamma}\frac{p_{1}}{p_{0}} +\frac{s_{1}}{c_{p}} \right] T_{0}.
\end{eqnarray}
Figures~\ref{fig:Fe}a, b, and c show meridional distributions of the enthalpy flux $F_{e}$ for the three extracted vorticity modes summed over $m=1-39$.
We note that all of these modes transport the thermal energy radially upward.
The $\zeta_{z}$-symmetric columnar convective modes transport about $65\%$ of what is required in the upper convection near the equator.
The $n=0$ Rossby modes and the mixed modes can transport about $8-9\%$ of the required thermal energy in the lower and upper convection zone, respectively, outside the tangential cylinder.

This is a striking result because the $n=0$ Rossby modes are believed to be quasi-toroidal and cannot contribute to the thermal energy transport in the linear theory for the simplified case with uniformly rotation, no turbulent diffusion, and adiabatic background stratification \citep[e.g.,][]{saio1982,damiani2020}.
We find that, under the influence of strong turbulent diffusion and superadiabatic background, the $n=0$ equatorial Rossby modes become partially convective especially at high $m$.

\subsection{Angular momentum transport}

These equatorial vorticity modes also transport the angular momentum.
Figures~\ref{fig:RS}a, b, and c show the Reynolds stresses $\rho_{0}\langle v_{r} v_{\phi} \rangle$ (upper panels) and $\rho_{0}\langle v_{\theta} v_{\phi} \rangle$ (lower panels) summed over all $m$ values for the extracted $\zeta_{z}$-symmetric columnar convective modes, $n=0$ equatorial Rossby modes, and the ``mixed Rossby modes'', respectively.
These terms are proportional to the radial and latitudinal components of the convective angular momentum flux.
For comparison, we show the total Reynolds stresses in our simulation (which include contributions from the other modes and small-scale convection) in Fig.~\ref{fig:RS}d.

As for radial transport of the angular momentum, the dominant contribution is from the $\zeta_{z}$-symmetric columnar convective modes that is about $8$ times bigger than those from the $n=0$ Rossby modes and the ``mixed Rossby modes''.
The radially-upward angular momentum flux by the columnar convective modes accounts for about $37\%$ of the total amount in the upper half of the convection zone near the equator.
The positive $\langle v_{r} v_{\phi} \rangle$ outside the tangential cylinder is a common feature of convection simulation in a strongly rotationally-constrained regime \citep[][]{gastine2013,fan2014,hotta2014b,matilsky2020}.

On the other hand, the angular momentum is preferentially transported equatorward in our simulation, as manifested by positive (negative) $\langle v_{\theta} v_{\phi} \rangle$ in the northern (southern) hemisphere in Fig.~\ref{fig:RS}d lower panel.
Our analysis reveals that both the $\zeta_{z}$-symmetric columnar convective modes and the $n=0$ Rossby modes contribute to this net equatorward angular momentum transport by about $30-40\%$ near the surface and at the base, respectively.
The ``mixed Rossby modes'' turn out to be rather insignificant for the net angular momentum transport in the latitudinal direction.

\section{Summary and Discussion} \label{sec:summary}

In this paper, we report a mode-by-mode analysis of the low-frequency equatorial vorticity modes in a fully-nonlinear simulation of solar-like rotating convection.
This study was motivated by the recent observational discovery of various types of inertial modes on the Sun \citep{loeptien2018,gizon2021} and the consequent theoretical study on these modes in a linear regime \citep[][]{bekki2021_l}.

Based on the equatorial velocity power spectra, we have successfully identified and characterized several types of equatorial modes in our simulation.
For each mode, eigenfunctions are extracted using the SVD method.
We have also carried out the linear eigenmode analysis with the simulated differential rotation included.
The computed linear dispersion relations and eigenfunctions are compared with the simulated power spectra and the extracted eigenfunctions.
Our work provides a technique for subsequent numerical studies of low-frequency inertial modes in nonlinear convection simulations.

We have successfully identified the $\zeta_{z}$-symmetric columnar convective modes, the $n=0$ equatorial Rossby modes, and the ``mixed Rossby modes''.
Table \ref{table:summary} summarizes the modes identified in this paper.
Although we have mainly focused on these equatorial modes in this paper, we have also checked that the high-latitude modes are found to exist in our simulation as well (see Appendix~\ref{appendix:C}).

Our major findings can be summarized as follows.
The north-south $\zeta_{z}$-symmetric columnar convective modes have the highest power in our simulation.
They originate primarily from the compressional $\beta$-effect near the surface and thus can be well characterized by the dispersion relation similar to that of \citet{glatzmaier1981}.
They transport a significant fraction of enthalpy upward and are the dominant term in the angular momentum transport near the equator.

Our analysis reveals that, at high $m$, the equatorial Rossby modes with no radial nodes ($n=0$) have eigenfunctions that  deviate from that of the uniformly-rotating and inviscid case ($r^{m} \sin^m\theta$).
As $m$ increases, we find that these modes are more and more confined near the base of the convection zone. 
We argue that this is due to the strong diffusion arising from the  turbulent convective motions on resolved scales, which breaks the radial force balance between the pressure gradient and the Coriolis force and drives radial flows \citep[][]{bekki2021_l}.
We find that these equatorial Rossby modes are the longest-lived modes in our simulation.

Mode mixing between the equatorial Rossby modes and the columnar convective modes is found in our nonlinear simulations, as predicted by the linear analysis of \citet[][]{bekki2021_l}.
The surface power spectrum of the north-south symmetric $v_{\theta}$ can be characterized by two oppositely-propagating modes that form a single well-defined power ridge across $m=0$.
The retrograde and prograde modes for $m>0$ can be identified as the $n=1$ equatorial Rossby modes and the north-south $\zeta_{z}$-antisymmetric columnar convective modes, respectively.
We call them ``mixed Rossby modes'' following \citet[][]{bekki2021_l}.
An analogy can be drawn between these ``mixed modes'' and the Yanai waves (mixed Rossby-gravity waves) where the retrograde-propagating Rossby waves and prograde-propagating gravity waves (Kelvin waves) are mixed with one another \citep[e.g.,][]{vallis2006}.
The existence of the ``mixed Rossby modes'' has important implications.
One of these follow from its frequency which is very close to that of the $n=0$ equatorial Rossby mode at $m \geq 5$ \citep[see Fig.~10 in][]{bekki2021_l}.
Therefore, it is possible that the observed Rossby modes on the Sun could be $n=1$ modes rather than $n=0$ modes as is typically assumed.

The nonlinear simulations contain a wealth of information about many more modes of oscillations in the inertial frequency range.
In fact, the analysis method reported in this paper can be used in the future to study the $m=1$ high-latitude inertial mode \citep[][]{gizon2021,bekki2021_l} and the $l=m+1$ high-frequency retrograde modes \citep[][]{hanson2022}.

\begin{acknowledgements}
We thank an anonymous referee for constructive comments.
Y. B. is enrolled in the International Max-Planck Research School for Solar System Science at the University of G\"ottingen (IMPRS). 
Y .B. also acknowledges a financial support from long-term scholarship program for degree-seeking graduate students abroad from the Japan Student Services Organization (JASSO).
We acknowledge a support from ERC Synergy Grant WHOLE SUN 810218 and the hospitality of the Institut Pascal in March 2022.
All the numerical computations were performed at GWDG and the Max-Planck supercomputer RZG in Garching. 
\end{acknowledgements}

\bibliographystyle{aa} 
\bibliography{ref} 

\begin{appendix} 

\section{Temporal evolution} \label{appendix:A}

\begin{figure}[]
\begin{center}
\includegraphics[width=0.98\linewidth]{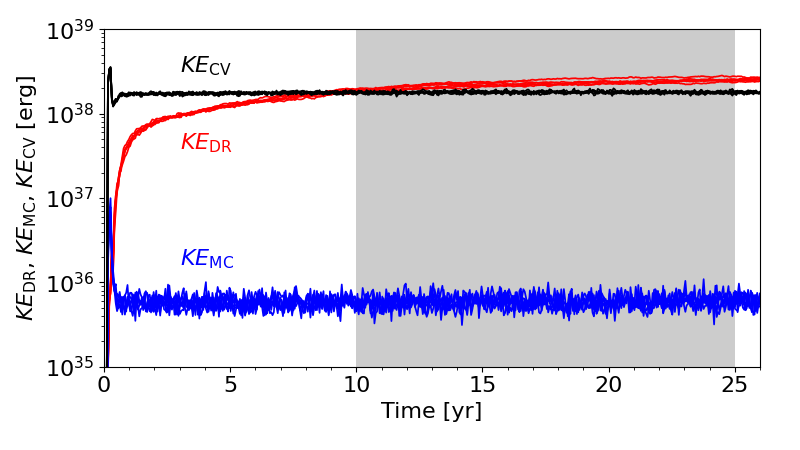}
\caption{
Temporal evolution of the volume-integrated kinetic energies;
(red) Kinetic energy of the differential rotation $\mathrm{KE_{DR}}$, (blue) kinetic energy of the meridional circulation $\mathrm{KE_{MC}}$, and (black) kinetic energy of the non-axisymmetric flows $\mathrm{KE_{CV}}$.
The results from the total 6 runs starting with different initial conditions are shown.
The grey shaded are denotes the duration which we use for our spectral analysis.
}
\label{fig:Ekin_time}
\end{center}
\end{figure}

In this appendix, we show the temporal evolution of the volume-integrated kinetic energies of the differential rotation $\mathrm{KE_{DR}}$, meridional circulation $\mathrm{KE_{MC}}$, and non-axisymmetric flows (turbulent convection and modes of oscillation) $\mathrm{KE_{CV}}$.
They are defined as
\begin{eqnarray}
\mathrm{KE}_\mathrm{DR} &=& \int_{V} \frac{\rho_{0}}{2} \langle v_{\phi} \rangle ^{2}  dV, \\
\mathrm{KE}_\mathrm{MC} &=& \int_{V} \frac{\rho_{0}}{2} [\langle v_{r} \rangle ^{2} + \langle v_{\theta} \rangle ^{2}]\  dV, \\
\mathrm{KE}_\mathrm{CV} &=& \int_{V} \frac{\rho_{0}}{2} \left[ (v_{r}-\langle v_{r} \rangle)^{2} + (v_{\theta}-\langle v_{\theta} \rangle)^{2}+ (v_{\phi}-\langle v_{\phi} \rangle)^{2}  \right] dV. \nonumber \\
\end{eqnarray}
Figure~\ref{fig:Ekin_time} shows their temporal evolution for 6 runs that are initiated from different random initial perturbations.
We find that the mean flow profiles are almost unaffected by the differences in the initial condition.
In all cases, the differential rotation reaches an almost statistically-stationary state at around $t \approx 10$ yr.
For our analysis, we use the $15$-year-long data after the differential rotation becomes quasi-stationary ($10$ yr $<t<25$ yr), as denoted by shaded grey area in Fig.~\ref{fig:Ekin_time}.

\section{Effective turbulent diffusivity} \label{appendix:B}

In this appendix, we give an estimate of the effective turbulent diffusivity in our nonlinear simulation $\nu_{\mathrm{eff}}$ due to the small-scale convective motions on resolved scales.
Using the first-order smoothing approximation (FOSA) \citep[e.g.,][]{moffatt1978}, we calculate $\nu_{\mathrm{eff}}$ as
\begin{eqnarray}
&& \nu_{\mathrm{eff}}=\frac{\tau_{c}}{3}v_{\mathrm{rms}}^{2}=\frac{v_{\mathrm{rms}} H_{p}}{3},
\end{eqnarray}
where the convective turnover time scale $\tau_{c}$ is given by $H_{p}/v_{\mathrm{rms}}$.
Here, the rms convective velocity is calculated at each height as
\begin{eqnarray}
&& v_{\mathrm{rms}}=\sqrt{ \langle (v_{r}-\langle v_{r} \rangle)^{2} + (v_{\theta}-\langle v_{\theta} \rangle)^{2}+ (v_{\phi}-\langle v_{\phi} \rangle)^{2} \rangle }.
\end{eqnarray}

\begin{figure}[]
\begin{center}
\includegraphics[width=0.93\linewidth]{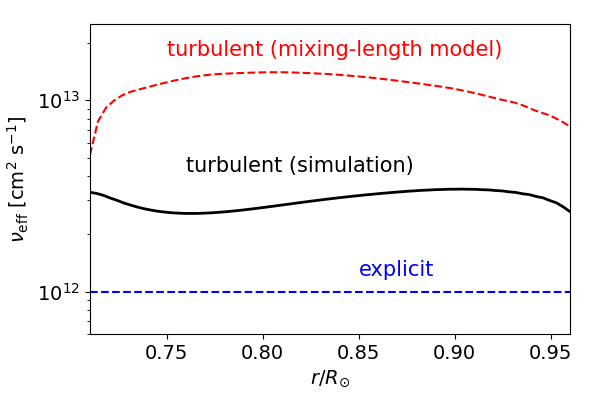}
\caption{
Radial profiles of the effective diffusivity due to small-scale convective motions $\nu_{\mathrm{eff}}$ (black solid).
For comparison, the explicit viscosity used in the simulation and the turbulent diffusivity estimated based on the mixing-length model \citep[][]{munoz2011} are also shown by blue and red dashed lines.
}
\label{fig:vrms_Ro_nueff}
\end{center}
\end{figure}

Figure~\ref{fig:vrms_Ro_nueff}c shows that $\nu_{\mathrm{eff}}$ is about $3\times 10^{12}$ cm$^{2}$~s$^{-1}$ throughout the convection zone.
This value is comparable but slightly bigger than the explicit viscosity value of $\nu=10^{12}$ cm$^{2}$~s$^{-1}$ which we use in the simulation.
This implies that the turbulent convective motions can have a impact on inertial modes by enhancing the effective viscous diffusivity.
Still, this enhanced value is smaller than the value estimated using a mixing-length model \citep[][]{munoz2011}.

\section{High-latitude inertial modes} \label{appendix:C}


\begin{figure*}[]
\begin{center}
\includegraphics[width=0.98\linewidth]{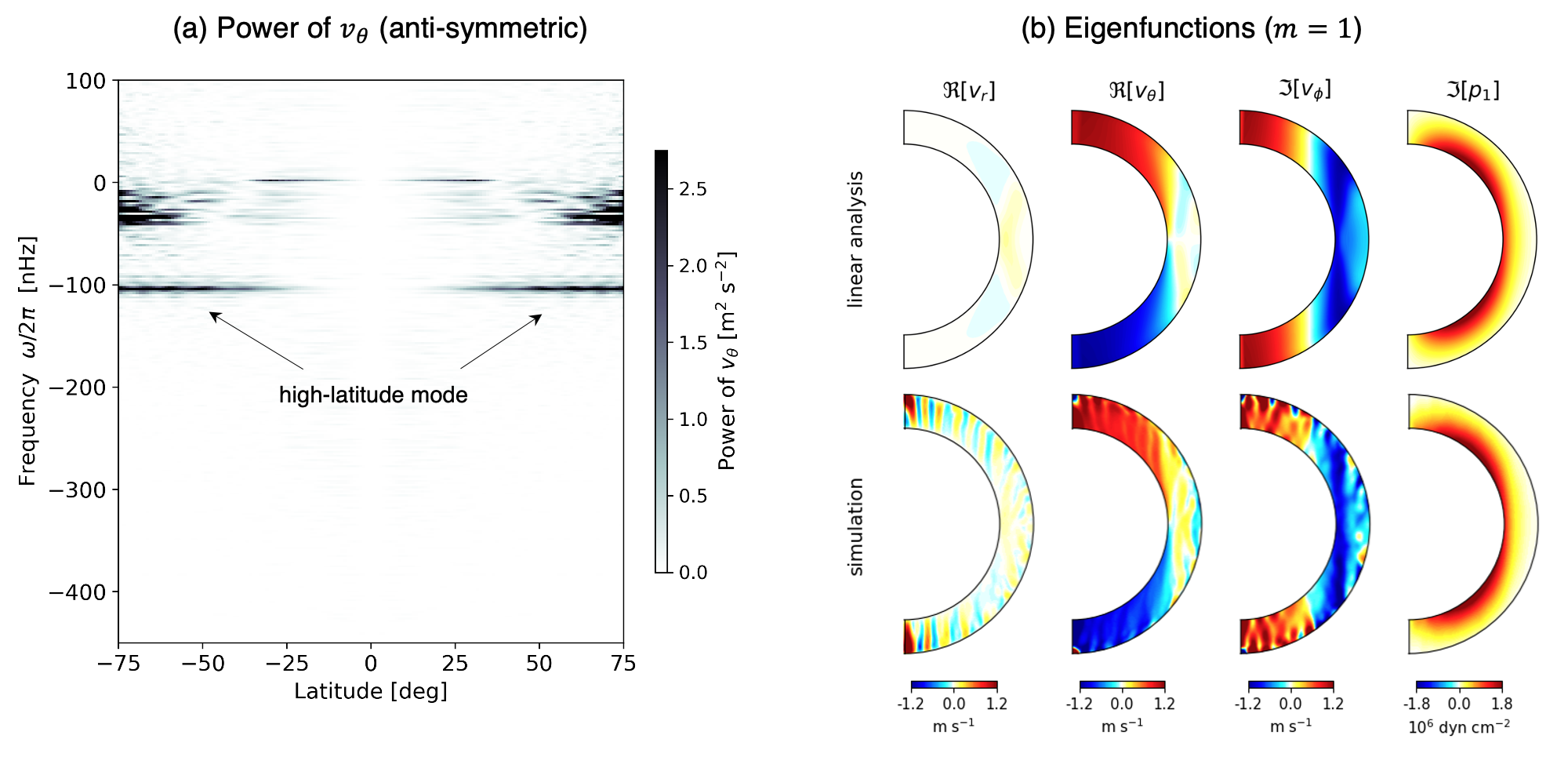}
\caption{
(a) Power spectrum of the north-south antisymmetric component of latitudinal velocity $v_{\theta}$ at the base of the convection zone $r=0.71R_{\odot}$ for $m=1$ as a function of latitude. 
The power associated with the high-latitude mode is denoted by black arrows at $\omega/2\pi\approx -105$ nHz. 
(b) Extracted eigenfunctions of the $m=1$ high-latitude mode (with north-south symmetric for $\zeta_{z}$). 
Lower and Upper panels show the eigenfunctions extracted from simulation and those of linear analysis, respectively. 
}
\label{fig:stpg}
\end{center}
\end{figure*}


\citet[][]{bekki2021_l} reported that there are retrograde-propagating $z$-vorticity modes that are predominantly confined inside the tangential cylinder.
They are called ``topographic'' Rossby modes because they originate from the ``topographic'' $\beta$-effect, i.e., spherical curvature effect of the lower and upper boundaries.
When seen at the surface, these modes are found only at high latitudes, and thus, are also called ``high-latitude'' modes \citep[][]{gizon2021}.

In this appendix, we show that the high-latitude modes, as well as the equatorial modes, can be unambiguously identified in our nonlinear rotating convection simulation.
We find that they exist predominantly at $m=1$. 
For $m=2$ and $3$, the power still exists but is substantially weaker.
Figure~\ref{fig:stpg}a shows the $m=1$ power spectrum of nouth-south antisymmetric component of $v_{\theta}$ as a function of latitude at the base of the convection zone $r=0.71R_{\odot}$.
The power of the high-latitude mode can be seen at $\omega/2\pi\approx -105$ nHz from middle to high latitudes.
The extracted eigenfunctions of this $m=1$ mode are shown on the lower panels of Fig.~\ref{fig:stpg}b and are compared with the results of linear analysis (upper panels).
The associated flow is dominantly $z$-vortical and is strongly confined inside the tangential cylinder.
The $z$-vorticity $\zeta_{z}$ is north-south symmetric across the equator.
The product of $\zeta_{z}$ and $p_{1}$ is negative throughout the domain, inferring that the mode is in geostrophic balance.

\begin{figure*}[]
\begin{center}
\includegraphics[width=0.98\linewidth]{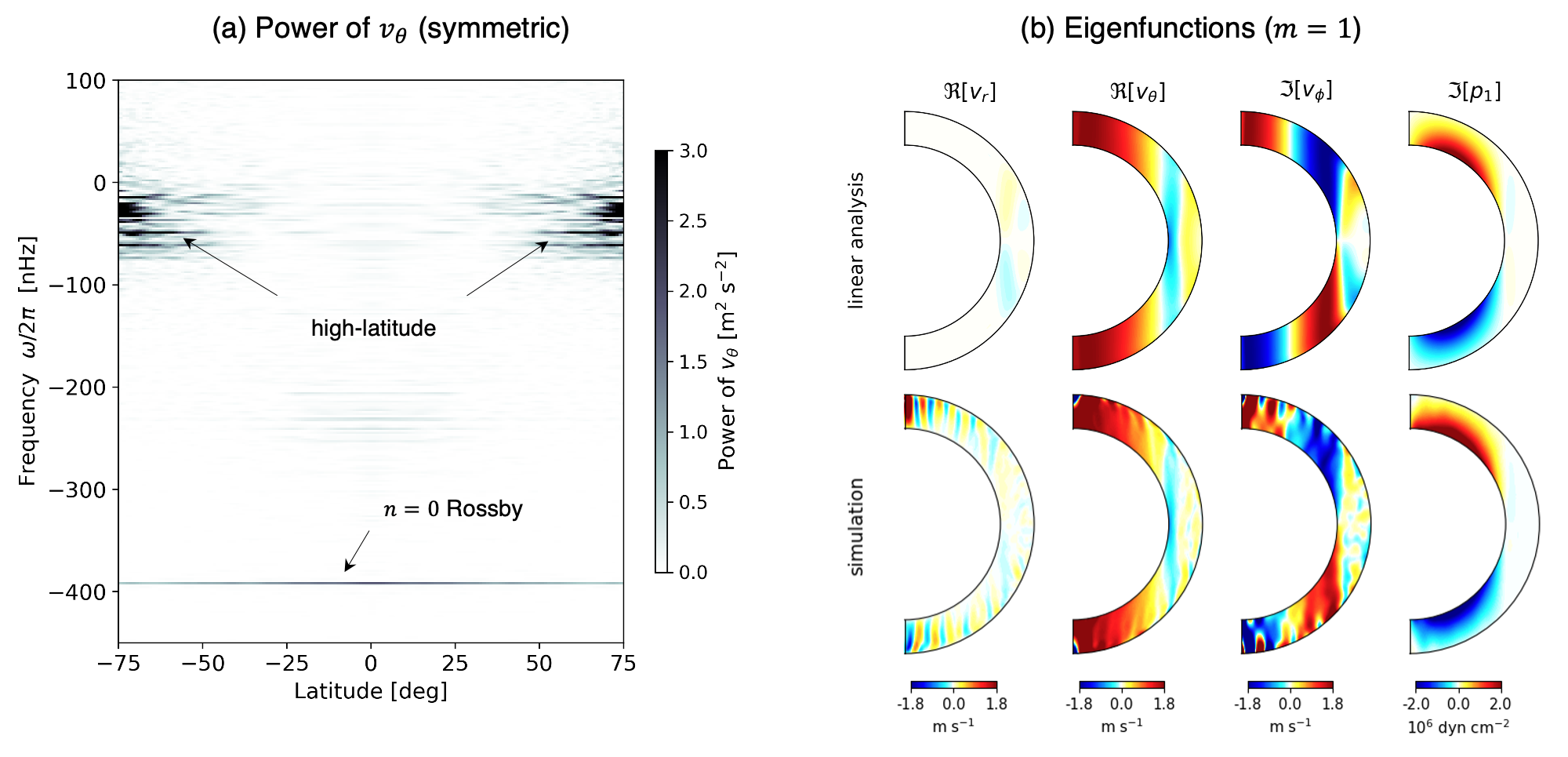}
\caption{
(a) Power spectrum of the north-south symmetric component of $v_{\theta}$ at the base of the convection zone $r=0.71R_{\odot}$ at $m=1$ as a function of latitude. 
The high-latitude mode's power exists at the frequency of about $-48$ nHz, 
whereas power of the $n=0$ equatorial Rossby mode can be seen at the frequency of $\approx -395$ nHz. 
(b) The same as Fig.~\ref{fig:stpg}b but for the mode with nouth-south antisymmetric for $\zeta_{z}$. 
}
\label{fig:astpg}
\end{center}
\end{figure*}

In addition to the $\zeta_{z}$-symmetric mode, we also identify the north-south $\zeta_{z}$-antisymmetric high-latitude mode in our simulation.
Figure~\ref{fig:astpg}a shows the same power spectrum as Fig.~\ref{fig:stpg}a but for the symmetric component of $v_{\theta}$ between the hemispheres.
The power of the high-latitude mode lies at $\omega/2\pi \approx -48$ nHz at high latitudes (above $\pm 50$ deg) whereas the $n=0$ equatorial Rossby mode is spotted at $\omega/2\pi \approx -395$ nHz at low latitudes.
Figure~\ref{fig:astpg}b shows the extracted eigenfunctions of $\zeta_{z}$-antisymmetric high-latitude mode at $m=1$.
Unlike the symmetric mode, $\zeta_{z}$ is dissected between the hemispheres.
However, there exists a latitudinal velocity at the equator, which can correlate the vortices between the hemispheres.

The horizontal velocity amplitudes of these high-latitude modes are $1-2$ m~s$^{-1}$ at $m=1$ (regardless of the north-south symmetries).
\citet[][]{bekki2021_l} reported that, when the latitudinal entropy variation is significant throughout the convection zone, these modes become baroclinically unstable and exhibit a spiralling pattern around the poles as observed on the Sun \citep[][]{hathaway2020,gizon2021}.
However, in the simulation reported here, the latitudinal entropy variation is too weak for the baroclinic instability to occur.
Therefore, the observed spiralling features are not reproduced.

\end{appendix}

\end{document}